# Systemic physiological cliff at menopause revealed by temporal deconvolution of 300 million lab tests: a multi-cohort retrospective study

Glen Pridham[1,†], Yoav Hayut[1,†], Noa Lavi-Shoseyov[2,3], Michal Neeman[2], Noa Hovav[4], Yoel Toledano[5], Uri Alon[1,*]

[1]Dept. Molecular Cell Biology, Weizmann Institute of Science; Rehovot, 7610001, Israel

[2]Dept. Immunology and Regenerative Biology, Weizmann Institute of Science; Rehovot, 7610001, Israel

[3]In vitro fertilisation unit, Kaplan Medical Center; Rehovot, 7610001, Israel

[4]Menopause Community Center, Clalit Health Care; Ness-Ziona 7407436, Israel

[5]Division of Maternal Fetal Medicine, Helen Schneider Women's Hospital, Rabin Medical Center; Petah Tikva, 4941492, Israel

[†]These authors contributed equally

[*]Corresponding author. Email address: urialonw@gmail.com

**Abstract**

**Menopause fundamentally reshapes female physiology, yet current understanding is limited by small longitudinal cohorts that characterize it as a gradual transition. Large-scale biomedical datasets remain underutilized because the age of the final menstrual period (FMP) is rarely recorded. Here, we present a computational framework that leverages cross-sectional data to reconstruct systemic physiology as a function of time relative to FMP. We adapted a deconvolution framework from astronomy to recover systemic biological trajectories by deconvolving the population distribution of FMP age from chronological data. Applying this to two national cohorts with 300 million laboratory tests from 1.3 million females, we transformed cross-sectional measurements into high-resolution timelines anchored to the FMP. Our analysis reveals a step-like physiological cliff at the FMP across endocrine, skeletal, hepatic, renal, inflammatory, and lipid systems. These discontinuities are absent in males and highly concordant across independent populations. We demonstrate that systemic dysregulation begins over a decade prior to FMP, significantly expanding the window for preventive intervention. Furthermore, hormone replacement therapy (HRT) appears to markedly attenuate these abrupt physiological shifts. These findings further support a systemic and quantifiable view of the menopausal transition and provide a generalizable strategy for recovering hidden biological trajectories from human datasets, applicable to other life stages such as puberty or disease progression.**

## Introduction

Menopause - the permanent cessation of menstruation - is an understudied transition in human health[1–5]. Understanding the dynamics of menopause, how physiological systems respond to these dynamics, and how early dysregulation begins, is therefore central to understanding human physiology and health.

Menopause is due to exhaustion of ovarian follicles, with sizable sex hormone fluctuations in the years before menopause, ultimately leading to a drop in estrogen and progesterone and rise in FSH, with variability between individuals[4]. Because sex hormone receptors are expressed in almost all human tissues, with effects on brain, vasculature, bone, liver, and adipose tissue[6–8], these hormonal dynamics reverberate beyond fertility to many physiological systems[3].

The menopause transition was illuminated by prospective cohorts such as the Study of Women's Health Across the Nation (SWAN)[5,9–12], showing that sex hormone changes begin about 7 years before menopause[5] and that bone loss accelerates in relation to menopause[11]. Such longitudinal studies are crucial due to their exquisite menopause documentation. However, longitudinal studies such as SWAN are limited in their number of participants, include only a small panel of laboratory tests, and follow subjects for only a few years prior to the final menstrual period (FMP). Larger omics screens[13] imply that menopause is a systemic inflection point, but the *de facto* use of age instead of menopause timing blurs the timeline of this onset, as sampling is done at an arbitrary time point for each participant relative to her FMP. Thus the field lacks both sample size and temporal resolution.

To address this, we develop a method to obtain detailed cross-sectional menopause dynamics from large medical datasets, and apply this to over 300 million laboratory tests from more than 1 million females. We build on a concept from astronomy that allows us to “deblur” the trajectories by deconvolving the known menopause age-of-onset distribution - similar to the way an astronomical image is sharpened by adjusting for the distortion caused by the telescope[14]. We resolve fine temporal dynamics as a function of time to menopause, revealing physiology-wide jumps at the FMP, which appear to be normalized by hormone replacement therapy, and are preceded by hormonal changes more than a decade before menopause.

## Results

### Analysis of two large datasets shows a jump at menopause in almost all physiological systems that is absent in males

We analyzed two large datasets (Fig 1a). The first is NHANES, a cross-sectional US health and aging study (1999-2021), with lab tests, health questions, exams and outcomes. The second is from health insurer Clalit that includes about half the Israeli population, including 1.3 million females aged 30-70 in the years 2004-2024. Lab tests in this anonymized all-comers dataset were analyzed cross-sectionally. For each lab test, we omitted data from individuals who took medications or had ICD9 codes that statistically affect that test ('healthy, no-med' dataset[15,16]). In both datasets we considered only nonpregnant individuals (demographics in Table 1).

We sought to study lab tests as a function of time from menopause. Standard cross-sectional analysis shows that the population means of test values change gradually with age and in females shows a higher slope at ages 40-60. As an example, the iron-related test ferritin is shown in Fig 1c.

The change around menopause is broadened due to the different menopause time of each individual, defined by FMP. To distill the effect of time from FMP, we assumed that each test value is an unknown function of time to menopause $f(t)$ where $t \equiv a - a_m$ is the difference between an individual's age $a$ and their age of FMP $a_m$. We developed an approach to determine $f(t)$, including a possible step-like change at menopause, by deconvolving the known FMP time distribution[17] from each lab test's cross-sectional trajectory. The FMP age distribution in NHANES has a median of 50 years and interquartile range (IQR) of 45-53 years (Fig 1c). We validated this method using simulated data and different forms of FMP distribution functions, including variations in average FMP age of 4 years, the observed range across human populations or major risk groups[18] (Supplementary Fig S1,S2).

To clarify, our algorithm does not predict the FMP time for each individual. Instead, we deduce the cross-sectional mean trajectory of each lab test as if we knew the age of FMP for each individual and aligned the tests by time from FMP.

We find that both datasets show jumps at menopause for almost all lab tests. For example, the ferritin test adjusted for time to menopause shows a jump (step-like change) of 0.6 standard deviations (STDs), where STD is the standard deviation of log-scaled test value (z-score) compared to a control population of 25-45 year old nonpregnant females (Fig 1c).

We show the dynamics in terms of time to menopause for 90 tests in Fig 2 (Supplementary Table 1). The YoGIen algorithm reveals jumps at menopause in females (Supplementary Fig S3), but not in males (Supplementary Fig S4). Females showed a significant jump in 52/90 lab tests (z test vs males, FDR corrected). Jump sizes ranged from 0.1 to 1.4 STDs.

We compared these to males, where we perform the same adjustment as females by deconvolving the same FMP distribution (apparent menopause time t=0 for males can be interpreted as age 50). Males did not show jumps around t=0, and instead their mean lab tests changed gradually (Fig 2a, blue lines).

In females, large jumps (≥0.5 STDs) are found in sex hormones, including estradiol (E2), anti-mullerian hormone (AMH) and follicle-stimulating hormone (FSH). These jumps are consistent with longitudinal data on E2 and FSH from SWAN (Supplementary Fig S5). More surprising is the prevalence of such jumps in almost all other physiological systems – liver tests, blood counts, blood chemistry, kidney functions, bone tests, lipids and osmolality (Fig 2b).

Despite differences in population, context and study design, the Clalit and NHANES datasets show strongly correlated jump sizes ($r=0.89$, $p=10^{-16}$). We also tested sensitivity to age range and use of only participants with recorded menopause status, and found excellent agreement ($r>0.95$, $p=10^{-16}$), Supplementary Fig S7.

Most of the observed jumps are for the worse in the sense that they are in the direction of increased disease risk. The main exceptions are tests related to iron, hemoglobin and red blood cells. The latter tests improve (move in the direction of less anemia) probably due to the cessation of menstrual bleeding[19].

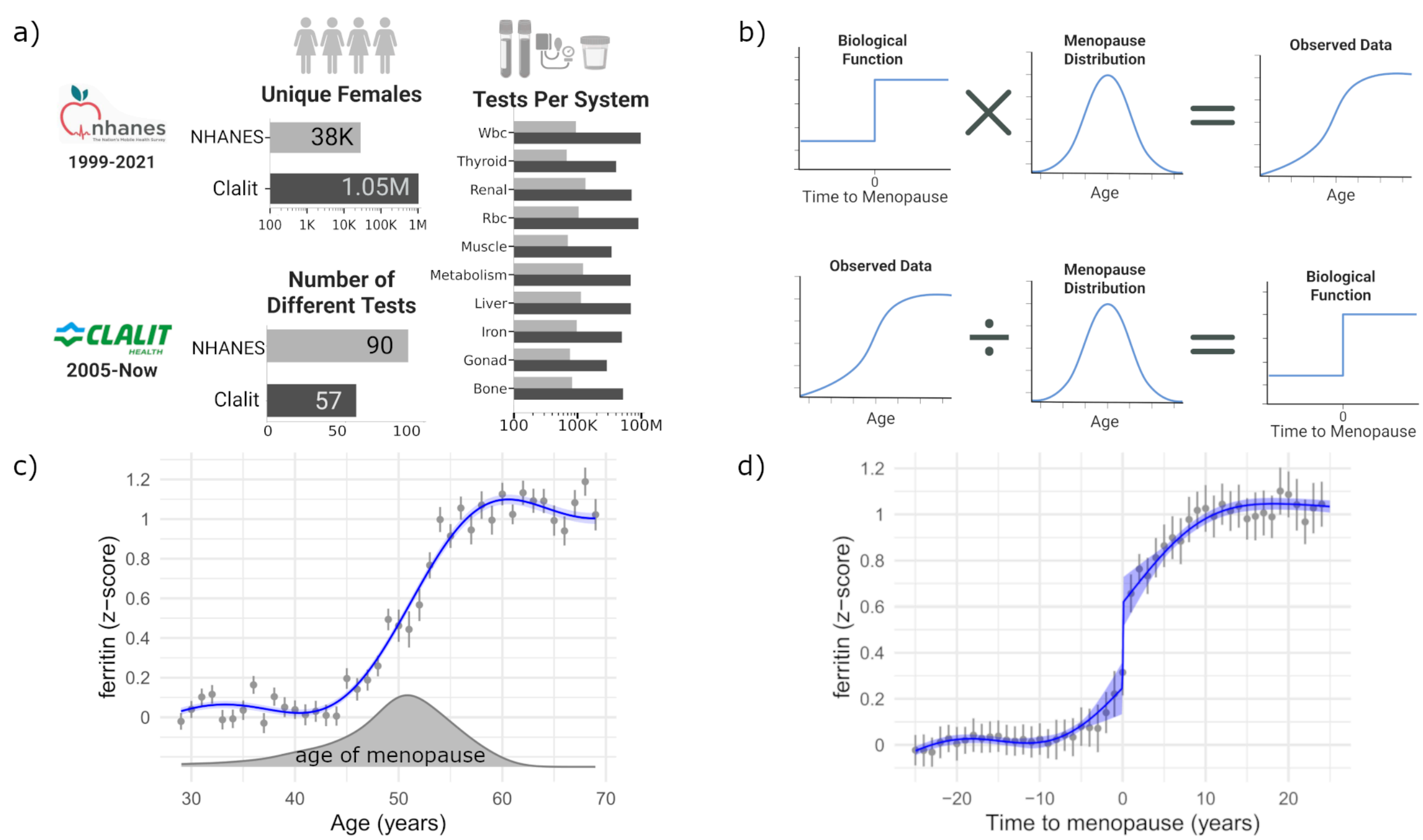

**Fig 1. Physiological tests as a function of time-from-menopause show a jump. a)** Overview of datasets (57/90 tests were in both studies) and **b)** deconvolution approach. **c)** representative trajectory of ferritin in females as a function of age, and the distribution of menopause ages (gray), from NHANES (smoothed with 1.5 year bandwidth). Blue line is a 10-knot spline. **d)** ferritin, adjusted for time to menopause using the deconvolution approach, shows a jump at menopause t=0 (z-score compared to non-pregnant reference group; from NHANES). Blue line is an average of flexible models weighted by cross-validation (see Methods). Error bars are standard errors of the mean.

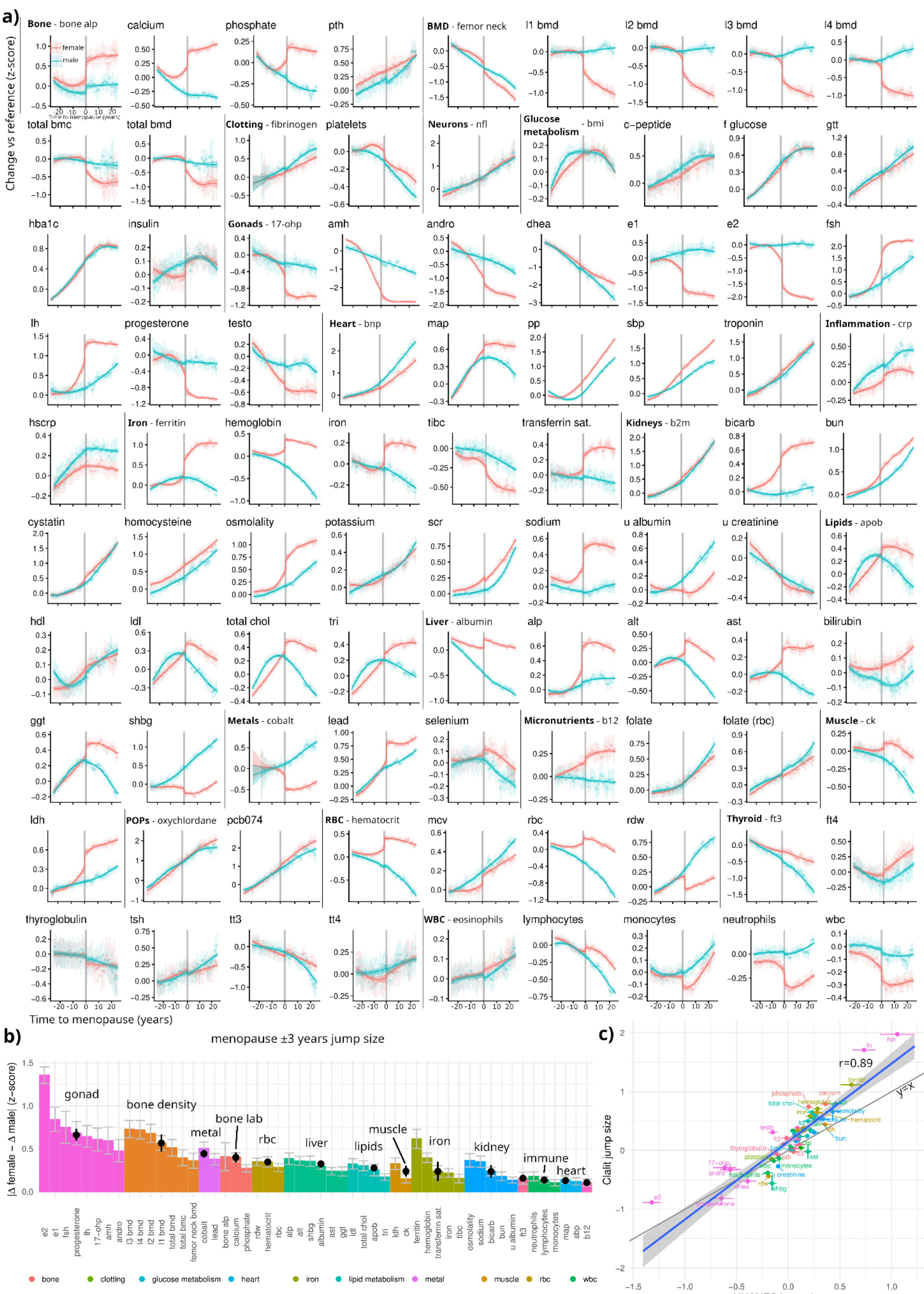

**Fig 2. Tests show jumps across systems in females but not males, and jump sizes are similar in the two datasets. a)** 90 lab tests from NHANES as a function of time to menopause, grouped by system. Females (red) often show a jump at t=0, whereas males (blue) do not (male data was deconvolved in the same way as females). Error bars are standard error of the mean. Lines are flexible model averages weighted by cross-validation (see Methods). **b)** NHANES jump sizes ordered by system, black points are median for each system (3-year change for females minus males; all $p < 0.05$, FDR corrected). **c)** Jump size in NHANES and Clalit datasets show high correlation ($r=0.89$, $p=10^{-16}$). Jump sizes are from the YoGlen algorithm (Methods).

**Hormone replacement therapy appears to reduce the physiological jumps at menopause**

We tested the effect of hormone replacement therapy (HRT, often called menopausal hormone therapy MHT) on the dynamics of lab tests. We used NHANES females with self-reported post-menopause status taking estrogen-containing drugs and/or those that answered the NHANES question *taking (estrogen, progestin, or combo) pill now*. Since this group amounted to only 9% of the participants, it may differ from the rest of the cohort in symptoms, access to healthcare and other factors. We therefore used two methods to adjust for economic and lifestyle characteristics: propensity score matching[20] and linear regression (Methods). Both methods produced concordant results ($r=0.90$, $p=10^{-16}$, Supplementary Fig S8). We specifically adjusted before vs after 2004, the year in which HRT usage dropped dramatically (Supplementary Fig S6).

Overall, 39/80 of the tests were different between HRT and non-HRT cohorts (Mann–Whitney U test, FDR corrected, min 10 propensity score matched individuals per test). This includes bone-dependent tests (BMD, bone alp, calcium), osmolality and sodium, liver tests (albumin, alp, alt, bilirubin and ggt) and renal tests (cystatin and homocysteine). Several additional tests show a trend for an effect that did not reach significance (9/80, $p < 0.1$, U test; Table S3).

We find that the size of most of the jumps at menopause is reduced in participants that reported current HRT usage (Fig 3a). The bigger the test jump at menopause, the larger the correction seen in HRT, Pearson's $r=0.76$ (95% CI: 0.65-0.84, $p=10^{-15}$; without iron and red blood cell (RBC) $r=0.81$, 95% CI: 0.64-0.85, $p=10^{-13}$). We also analyzed tests by system, to find that in most systems that showed a significant jump, notably excepting c-reactive protein (CRP) and iron/RBC, the absolute jump magnitude was significantly lower in the HRT cohort (Fig 3b) especially in the bone, liver and kidney/osmolality systems.

We compared the NHANES findings to Clalit. We analyzed females from Clalit who were prescribed HRT drugs (angeliq, EVIANA) at least twice (n=9103). We used a longitudinal design, comparing these females before, during and after HRT use. We compared to control females from the same clinic, age and birth year. We find moderately-strong concordance between the HRT effect sizes with NHANES ($r=0.63$, $p=0.005$) (Methods).

We conclude that HRT appears to reduce the jump-like change at menopause, making the post-menopause traces resemble a smoother continuation of the premenopausal traces.

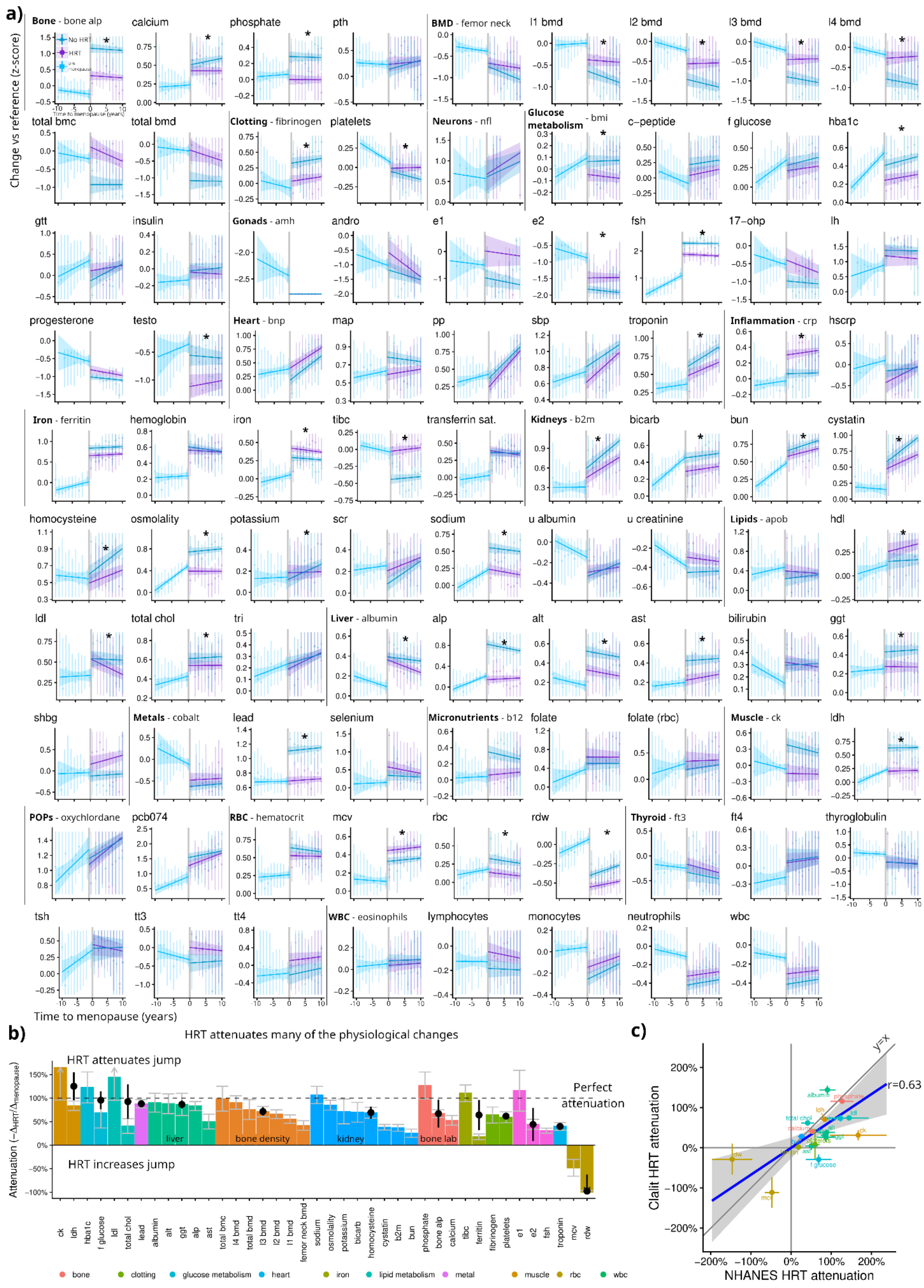

**Fig 3. Hormone replacement therapy (HRT) appears to reduce jumps in many systems. a)** NHANES lab tests as a function of time to menopause comparing participants with current HRT use (purple) to propensity score matched controls (blue; median n = 378 IQR: 114-710 with test and taking HRT, see Supplementary Table S3); matched premenopause participants also in blue. Star marks significant HRT effects (FDR corrected paired Mann–Whitney U test $p < 0.05$). **b)** Jump attenuation defined as the ratio of jumps in HRT versus control, (HRT-pre)/(control-pre), shown are significantly changed tests (FDR corrected unpaired z-test $p < 0.05$). 100% attenuation indicates complete abolition of the jump at the final menstrual period (FMP), negative values means a larger jump in the HRT cohort. Tests are grouped by system, black points are the median±mean absolute deviation of the system. All lab tests have at least 10 measurements for each group (HRT now, no HRT, and premenopause). **c)** NHANES and Clalit show moderate-strong agreement on HRT attenuation for the overlapping variables from b), ($r=0.63$, $p = 0.005$; including only Clalit variables with >200 individuals, Methods).

**Sex hormone dysregulation begins more than a decade before the FMP**

Three major sex hormones showed a marked change in dynamics long before the FMP (Fig 4). AMH begins an accelerated drop at about 15 years before menopause in its mean cross-sectional trajectory (Fig 4a,d). The gonadotropin FSH begins an accelerated rise about 14 $\pm$1 years before menopause. Estradiol (E2) begins an accelerated drop about 12$\pm$1 years before menopause. Change point analysis of Clalit deconvolved trajectories shows nearly identical times for the dysregulation of E2 and FSH (purple lines, Fig 4). These E2 and FSH changes are about 7 years earlier than documented in previous studies[5] (Fig 4).

The variation (STD) of hormone levels rises in the decade before the FMP (Fig 4 g-i). This is consistent with sex hormone fluctuations[21] and cycle irregularity[22] in the late menopause transition.

Mean E2 increases from 25 to ~12.5 years before menopause (Fig 4c) and its variation decreases (Fig 4i). This is consistent with reports that cycles become more regular and the follicular phase shortens from ages 25-40 (~25 to 10 years before menopause)[22].

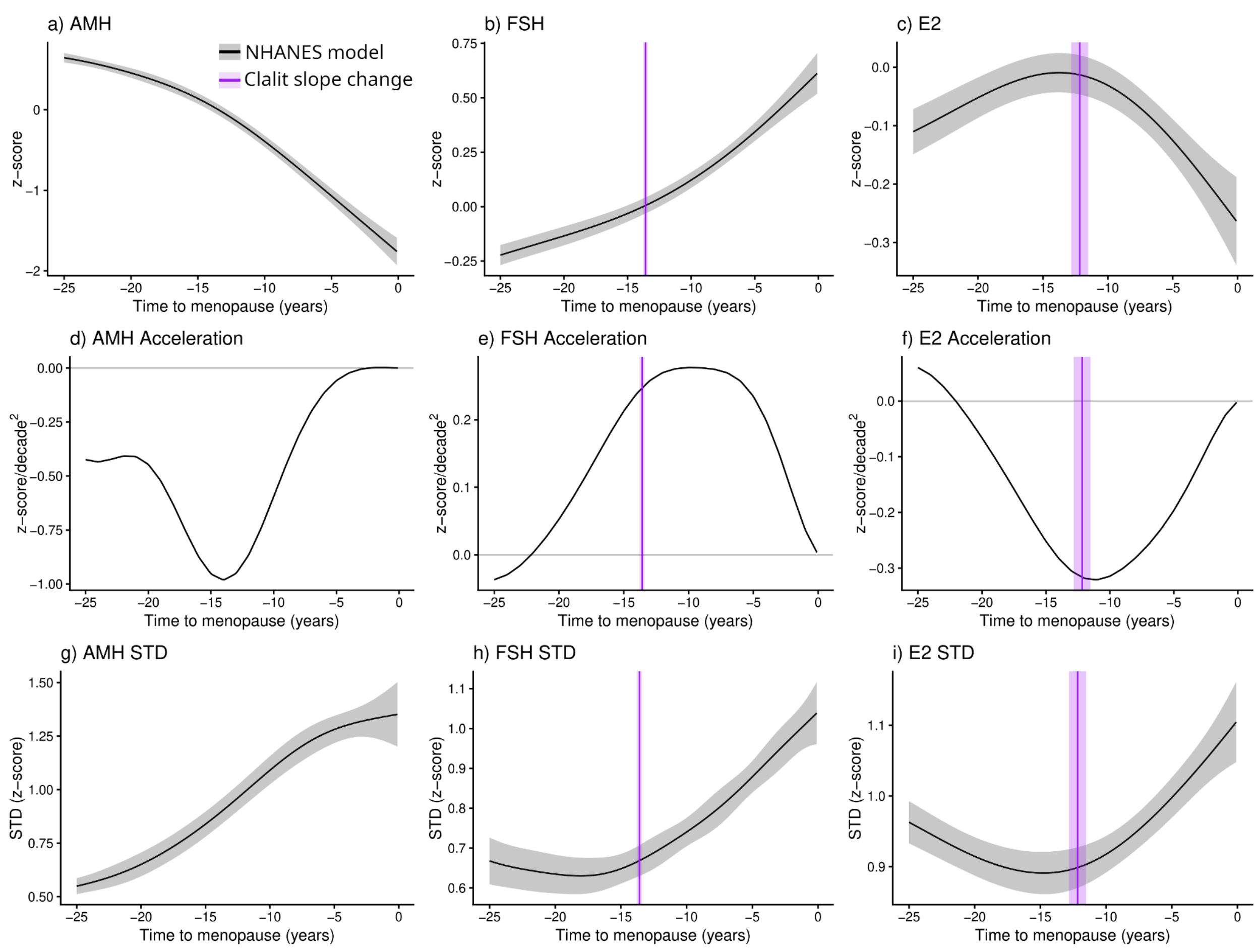

**Fig 4. Sex hormones show accelerated change more than a decade before menopause.** **a-c)** Cross-sectional means in the 25 years before the FMP from NHANES females who are self-reported non-menopausal (n=2406, 3035 and 4621, for anti-mullerian hormone, AMH, follicle-stimulating hormone, FSH and estradiol, E2, respectively). Purple lines are slope change points from the Clalit dataset (Methods) (Clalit AMH had insufficient data). **d-f)** acceleration (second derivative by time, also called curvature) of hormone traces, changepoints are located at the start of large accelerations (positive or negative), **g-h)** standard deviation of hormones. All measures are based on z-scores of log test values.

## Prevalence of health deficits show increased slope after menopause

Since most physiological lab tests show a jump-like shift at menopause, we used our deconvolution approach to ask about clinical outcomes as a function of time to menopause using data from NHANES. These health deficits include diagnoses of diabetes, osteoporosis, liver disease, heart disease, and symptoms such as trouble falling asleep.

Most of the deficits rise in prevalence after menopause. Several deficits also show a small jump at the FMP (Fig 5). Osteoporosis has a rise in slope at the FMP (Fig 5a), strongly outpacing males, consistent with the large effects of the menopause transition on bone tests. Cardiovascular outcomes (heart attack, stroke) also accelerate after menopause (Fig 5b).

Anemia was the only major clinical outcome that showed a drop in prevalence at menopause (Fig 5c) – consistent with the stop of menstruation leading to improved iron and RBC lab tests (difference in log-odds ratio vs males at FMP$\pm$3 years: -0.63$\pm$0.14, $p<10^{-3}$, FDR adjusted).

Thyroid and kidney outcomes are weakly affected by menopause, consistent with the relatively minor effect of menopause on renal function tests (creatinine, cystatin) and thyroid tests (Fig 2).

Whereas the prevalence of most clinical outcomes rises after menopause and stays high, depression scale ratings (PHQ9) showed a transient rise on average (Fig 5d). Moderate and severe depression rates (PHQ9 above 10) in post-menopausal females with self-reported menopause timing were significantly elevated in the year after menopause (46/236=19%) compared with the subsequent 24 years (698/6495=11%; Fisher test, $p < 10^{-4}$). Several symptoms and behaviors also rise transiently after menopause and then decay to the slower aging-like behavior seen in males. This includes difficulty falling asleep and insufficient sleep (Fig 5e).

Males also show a change in slope of prevalence curves around age 50 with no detectable jump. The gap between males and females widens for several clinical outcomes, with bone-related outcomes and thyroid showing increasingly higher prevalence in females. The gap between males and females narrows for anemia, liver disease, and several sleep deficits after menopause.

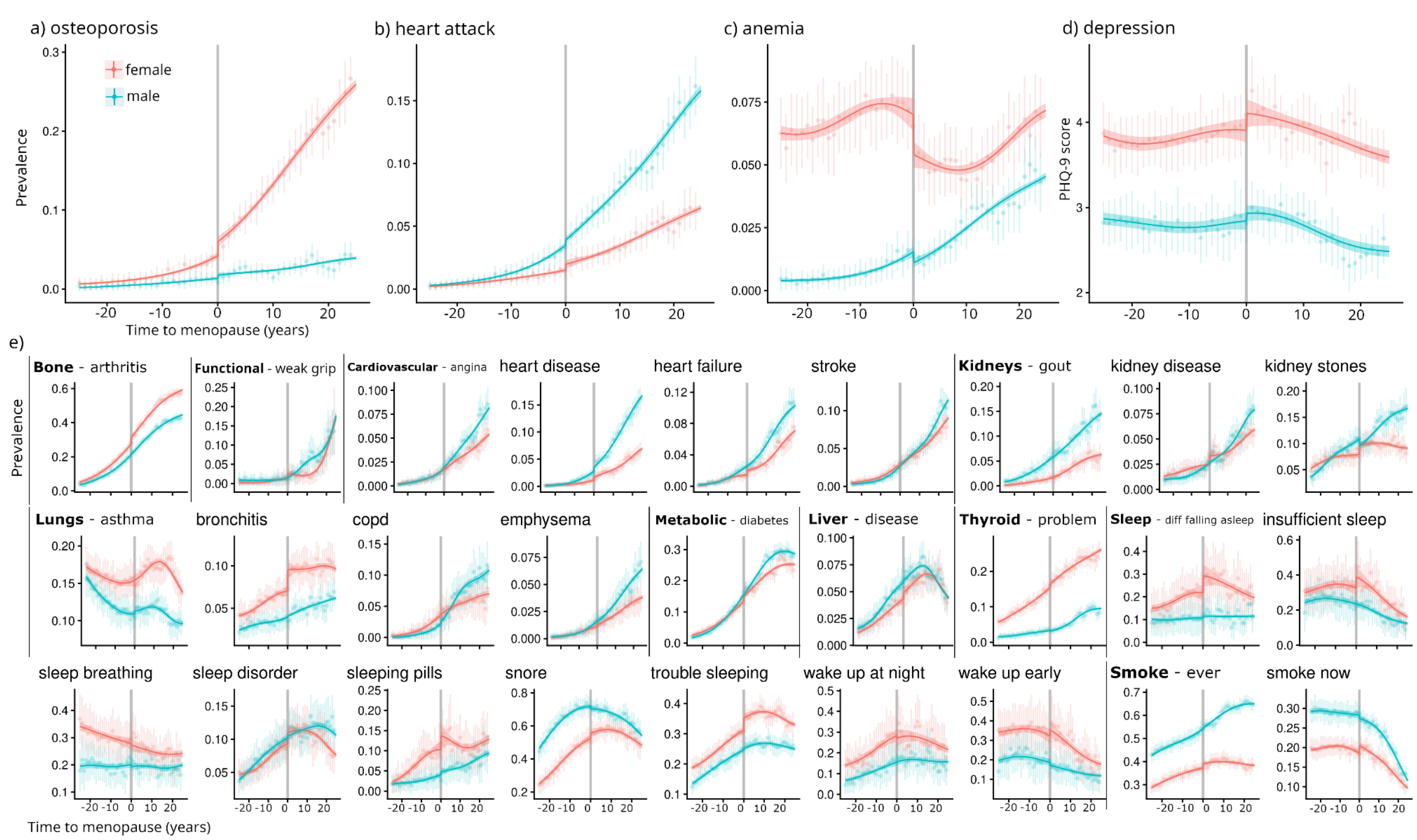

**Fig 5. Health outcomes and deficits generally show increased slope after menopause.** Prevalence in NHANES of **a)** osteoporosis, **b)** heart attack, **c)** anemia, **d)** depression score, **e)** 31 outcomes and deficits organized by system. Females (red) and males (blue). Lines are flexible model averages, weighted by cross-validation (see Methods). Details in Supplementary Table S4.

## Discussion

We analyzed two large datasets with over 300 million laboratory tests, indicating that menopause is characterized by a "physiological cliff," defined as abrupt, step-like changes across nearly all physiological systems. We introduce a temporal deconvolution framework that uses cross-sectional data to generate a high-resolution timeline anchored to the FMP (final menstrual period). This can detect systemic shifts that are typically obscured by chronological age. The observed jumps at the FMP are concordant between the United States (NHANES) and Israel (Clalit) cohorts. The changes are most pronounced in systems with high estrogen sensitivity, such as bone metabolism, liver tests, lipid handling and osmolality. While prior longitudinal studies, like SWAN, have established bone loss and specific hormonal surges, the scale of this study, covering 90 types of tests, demonstrates that this transition is sharp and systemic.

The magnitude of the jump in each test corresponds in part to the differential sensitivities of physiological systems to estrogen (Fig 2b). The large change of upstream hormones like FSH is due to their exquisite estrogen sensitivity via negative feedback in the hypothalamic-pituitary-ovarian axis. Bone is very sensitive to the protective effects of estrogen[24], followed by the liver[8] which is a hub for lipid metabolism. Osmolality is regulated by antidiuretic (ADH) hormone (also called arginine vasopressin), whose sensitivity is governed by estrogen[25]. Toxin levels such as lead also show a jump in menopause, probably due to the bone storage of lead[26]. Several tests showed no detectable jumps at the FMP, including free thyroid hormones[27], PTH[28], folate[29] and fibrinogen[30]. These systems are known to be weakly, indirectly, or debatably affected by estrogen and menopause. The jumps are preceded by accelerated changes in sex hormones more than a decade before menopause, earlier than the 3-7 year window traditionally ascribed[21].

Notably, we find that males lack these abrupt physiological shifts, suggesting that the cliff is a consequence of the menopausal transition rather than generalized aging.

These lab test shifts are associated with a rise in risk for various clinical outcomes, which manifest in our analysis as an increase in the slope of incidence at the FMP. Thus, prevalence is akin to an integral over time of the lab test trajectory. For instance, osteoporosis shows an increase in slope after menopause that strongly outpaces that of age-matched males, consistent with the large effects observed in bone-related laboratory tests. Cardiovascular outcomes, such as heart attack and stroke, also demonstrate an increased risk slope following the transition. Furthermore, we observed a transient rise in average depression scale ratings and sleep-related deficits around the time of the FMP, which then tend to decay back toward the slower aging patterns seen in males. We note that although the average rise in depression scores is transient[12], a subset of females are known to show sustained depression post-menopause[23].

Collectively, these findings suggest that the abrupt shifts in laboratory markers serve as a systemic inflection point for the accumulation of clinical health deficits.

Hormone replacement therapy, often referred to as menopausal hormone therapy (MHT), appeared to markedly attenuate the physiological shifts. In females reporting current HRT use, post-menopausal trajectories resembled smoother continuations of pre-menopausal states across various markers, including liver, renal, and bone tests. While HRT appeared to normalize most systems, it was associated with elevated CRP, an inflammation marker. This is consistent with studies that show that HRT estrogen stimulates hepatic CRP production in oral but not transdermal (patch) treatment due to a first pass hepatic effect[31,32]. In general HRT is anti-inflammatory[33,34].

This study has several important limitations. It relies on retrospective cross-sectional data and self-reported menopause status. While deconvolution sharpens the timeline, it cannot replace the granularity of a prospective longitudinal study with detailed menopausal tracking that follows the same individuals through the transition. Adjusting for sociodemographics cannot entirely eliminate healthy user bias among HRT users, because those who choose or have access to HRT may differ intrinsically from non-users in ways not captured by medical records or demographic data. The effects of HRT need to be tested by prospective randomized trials. The accuracy of our timelines also depends on the assumed distribution of FMP age, and while sensitivity analyses showed negligible effects to shifting this distribution, more accurate menopause tracking is warranted. Finally, while we controlled for many drugs, the lack of comprehensive data on hormonal IUDs or endometrial ablations in some cohorts could mask the true FMP for a subset of participants.

In conclusion, these findings further support a systemic and quantifiable view of the menopausal transition. We observed a physiological cliff at menopause across systems, which appears to be attenuated by hormone replacement therapy. This study extends the time frame for detectable changes and potential preventative care to almost a decade before menopause. More broadly, the present deconvolution approach can sharpen biological transitions from noisy, registry-scale data, a strategy transferable to other life-course events such as puberty, pregnancy and disease onset.

## Methods

### Data and cohorts

We analyzed two datasets originating from the USA (NHANES) and Israel (Clalit). NHANES is a cross-sectional health study that has been collected every 2 years from 1999-2022 that includes lab tests, health questions, exams and outcomes[35]. Clalit is a large electronic medical record (all-comers) longitudinal dataset that includes about half the Israeli population[15]. We analyzed healthy, no meds lab test data that excluded confounding effects of medications and disease[15], resulting in 1.3 million eligible females aged 30-70 from years 2004-2024.

**Eligibility criteria**

**Clalit**: We analyzed all comers aged 30-70 from years 2004-2024. We excluded pregnant females (N=173,000).

**NHANES**: We analyzed 12 biennial waves 1999-2022. We excluded pregnant females (N=1326). For HRT comparison (Figure 3) we only analyzed females from survey years 1999-2019 (2022 had no HRT data due to Covid), with l covariates needed for matching: menopause status, age, poverty income ratio, smoker status, insurance status, diabetes status, ethnicity and education, with final N=738 HRT users.

**Variable definitions**

**Menopause status (NHANES only)**

We included all non-pregnant females regardless of the reason for menopause (in sensitivity analysis we restricted to natural menopause at typical ages, showing highly concordant results with Pearson's r=0.97, Supplementary Fig S7). Menopause status (females only) was determined from the clinical definition of 12 months without period[2] together with supplemental information to confirm status. We used two questions, the first is: have you "had regular/at least one period in the past 12 months?". The second follow-up question is "what is the reason (for no periods)?", where we require the individual to specify "menopause" as the reason. Individuals that reported having periods were defined as non-menopausal (RHQ030/RHQ031: regular periods/at least one period in the last 12 months). Individuals were further classified as follows. Those self-reporting not having periods were classified as non-menopausal if self-reported pregnant (this year or previous) or breastfeeding, whereas individuals who self-reported menopause were classified as menopausal (RHQ040/RHD042/RHD043: reason not having regular periods); individuals that reported any other reason for not having periods were classified as unknown (NA) menopause status. RHD042 did not differentiate menopause from hysterectomy as the reason for not having regular periods but these cases were corrected afterwards as follows. We reclassified all females who had a hysterectomy that reported having either a single ovary or no ovaries removed as unknown (NA) menopause age and status, since the hysterectomy would prevent them from having periods (RHQ300/RHQ305/RHQ310/RHQ320); females who reported their menopause age before their hysterectomy age (RHQ291) were still included. We included participants taking oral contraceptives. Some of these individuals may not have periods in which case they were classified as unknown menopausal status. Finally, some individuals with indeterminate menopause status were classified according to age as follows: non-responders to regular periods (RHQ030/RHQ031) under the age of 30 were classified as non-menopausal, and responders over the age of 60 who reported not having regular periods were classified as menopausal (justified by the observed statistics that showed 98.3% of females are non-menopausal at age 30 and only 0.9% of females were non-menopausal at age 60). Self-reported age of menopause is top-coded at 60 in NHANES, that was extrapolated using survival modelling[36] as described in Methods subsection YoGlen Method 1. Clalit did not have menstrual data so menstrual status was not used for that dataset. Males and the 26.8% of females with unknown menopause status were imputed with a menopause age using the YoGlen algorithm described later in the methods section. We performed sensitivity analysis to consider the effects of unknown menopause status, hysterectomy and early/late menopause in Supplementary Fig S7, that shows nearly perfect concordance of results, r=0.97, between a restricted age range of 45-53 years old with known menopause status and our reported FMP jump sizes.

**Hormone replacement therapy (HRT)**

**NHANES:** We considered females as currently taking HRT if they had known post-menopause status and either (i) self-reported taking HRT or (ii) reported taking any drugs containing estrogens (full list in Supplementary Table S5). For (i) we gated NHANES using RHQ540 (ever used female hormones excluding birth control): any participant reporting "no" was classified as no HRT. Affirmative to any of the following was classified as yes HRT: RHQ558 (taking estrogen-only pills now; 2.9% of females), RHQ566 (taking progestin-only pills now; 0.2% of females), or RHQ574 (taking combo pill now; 1.1% of females). After 2012 NHANES stopped asking specific questions regarding HRT usage. For (ii) we used the RQQ_RX prescription drug files from each year except 2022 (not reported). We included the following types of estrogen-containing drug classes as HRT: all "estradiol", estropipate, mestranol, all "estrogen", as well as 9 individuals using progesterone-only HRT.

**Clalit:** In Clalit, menopausal status is not recorded. We therefore defined HRT users as females aged 45–55 who were prescribed at least twice the HRT drugs Angeliq (drospirenone 2 mg / estradiol hemihydrate 1 mg) or Eviana (estradiol hemihydrate 0.5 mg / norethisterone acetate 0.1 mg). We matched these individual users with controls from the same clinic, age and birth year. For each laboratory test and each group, we defined the menopause change Δ as the mean of the post-menopausal window (ages 54–60) minus the corresponding mean in the pre-menopausal window (ages 42–48). ΔHRT is the mean menopause change for HRT patients, Δcontrol is the mean menopause change for the control group. HRT attenuation is defined as the relative reduction in the menopause change of HRT users compared with their matched controls: $1 - (\Delta HRT / \Delta control)$.

**Variables and Outcomes**

We analyzed 90 lab blood tests and 31 outcomes for their time dependence with respect to inferred time of FMP. We group by systems as described in Supplementary Table S1,S4. Blood tests describe bone, liver, kidney, immune and other systems. Outcomes include signs, symptoms, and diagnoses.

**Preprocessing**

**Clalit and NHANES**: All lab values were log scaled and standardized to sex-specific mean and standard deviation using the control reference age range of 25-45 (datasets were standardized separately). An outlier cutoff was applied at $\pm\ 5$ STDs (<0.3% of datapoints). Log lab value distributions resembled normal distributions.

**Clalit**: Lab values were filtered to exclude measurements from non-healthy individuals and from individuals taking medications that affect each specific lab test. To do so, we used the dataset described by Cohen et al[15], which excluded 131 chronic conditions and 5,223 drug-test pairs.

Pregnancies were excluded separately. BMI was linked cross-sectionally for the healthy population using the most recent measurement within the study period, and reported as median [IQR]. Bootstrapping on Clalit was performed by grouping individuals by year of test date and sampling the 20 years with replacement. Test values of zero were removed (<0.3% of datapoints, Supplementary Table S2).

### The YoGlen Algorithm

We estimate the unknown age of FMP using a latent variable model[37] with appropriate approximations. Let $p$ denote an arbitrary probability density function, $a_m$ is the unknown age of FMP, $m$ is the binary menopause status variable (available in NHANES but not Clalit), $a$ is the known age, and $\vec{y}$ is a vector of known lab test values. The two results we derive are (1) how to estimate the distribution of the unknown age of FMP of each individual, $p(a_m|\vec{y}, a)$using the observed data, and (2) how to estimate mean cross-sectional lab trajectories as a function of the unknown time to FMP, $E[y_j|a]$. Jump size was estimated using resampled regression model predictions. For Method 1 we used 10 repeat, 10 fold cross-validation. For Method 2 we used bootstrapping of $E[y_j|a]$.

### YoGlen Method 1 for NHANES data

To estimate the unknown FMP time, $p(a_m|\vec{y}, a)$ we use Bayes' theorem $p(a_m|\vec{y}, a, m) \propto p(a_m|a, m)p(\vec{y}|a_m, a, m)$. We observed based on comparison to SWAN data (Supplementary Fig S5) that a good approximation was to use only the lowest order term $p(a_m|\vec{y}, a, m) \approx p(a_m|a, m)$. We used draws from $p(a_m|a, m)$ to impute the unknown $a_m$ where $p(a_m)$ is the reference menopause age distribution from NHANES (Fig 1), estimated by modelling the log-hazard as a 10-knot natural spline[36]. Females with double oophorectomies (12%) were drawn separately using their own distribution. For females where menopause status $m$ is known (73.2%), conditioning on $a$ and $m$ restricts the permitted $a_m$ to be before current age if non-menopausal and after current age if menopausal. We used a discrete grid to consider all possible $a_m$ from 20-70 at 1 year resolution. Once the $p(a_m|a, m)$ grid is estimated for each individual, we use it to draw 15 possible $a_m$ for each individual, creating 15 completed datasets that are analyzed in parallel and then the results are pooled using Rubin's rules[38] to account for the uncertainty in $a_m$.

In NHANES, most (89%) of the self-reported post-menopausal females also included their age of menopause. Prior research indicates that self-reported age of menopause is accurate to about 1 year[39], and hence age of FMP for these participants was estimated assuming $p(a_m)$ was a normally distributed true age of menopause with mean equal to their reported value and a 1 year standard deviation. Validation by comparison to longitudinal data from SWAN demonstrates that YoGlen fits well to the population mean for the key variables available in SWAN, FSH and E2, when only age and menstrual status are included (Supplementary Fig S5).

Inclusion of FSH and E2 into the model tended to fit worse (not shown), ostensibly because of the strong heteroscedasticity due to fluctuations in sex hormones that occurs during perimenopause[21,40] and suddenness of the transition. Therefore for NHANES the overall population trends were estimated using only menstrual status and age.

**YoGlen Method 2 for Clalit data**
For Clalit we did not have access to menstrual status data, precluding using YoGlen Method 1, so we instead used a regression model to deconvolve each lab test to convert the age-dependent cross-sectional means into time-since-FMP dependent dynamics, effectively using $p(a_m|y_j, a, m) \propto p(a_m|a)p(y_j|a_m, a)$. We modelled each lab test as piecewise linear,

$$f(t|a) \equiv \beta_0 + \alpha_0\ t + (\beta_m + \alpha_m t) \cdot I(t > 0) + \sum_{i=1}^{p}\left(\alpha_i \cdot (t - t_i) \cdot I(t > t_i)\right),$$

where $t \equiv a - a_m$ is the time to menopause, $I$ is the indicator function (1 when true, 0 otherwise), p is the number of changepoints (p=1 for Fig 4 and p=0 otherwise), and $\alpha_i$ and $\beta_i$ are the regression model parameters. The $\beta_m$ parameter permits a jump at FMP (t=0). Averaging this equation over age of menopause $a_m$ yields the cross-section mean as function of age that was fitted using least squares regression on the age-binned average values,

$$E[y|a] = \beta_0 + \alpha_0 E[t|a] + \beta_m C_m(a) + \alpha_m \int_0^a C_m(a_m)da_m + \sum_{i=1}^{p}\alpha_i \int_0^{max(0,a-t_i)} C_m(a_m)da_m.$$

where $E[t|a] = a - E[a_m]$, $E[a_m]$, and $C_m$ is the cumulative density of the FMP, these values were computed numerically from a published menopause age distribution from 306,068 healthy females[17]. For the changepoint results (Fig 4) we considered all possible $t_1 < 0$ (pre-menopause) and selected the best location for the changepoint ($t_1$) based on the lowest Bayesian Information Criterion.

**Propensity score matching (PSM)** was used to match HRT recipients after menopause to controls both before and after menopause (1:1 matching ratio). Propensity score matches each individual to similar controls using logistic regression on a set of covariates to determine similarity[20]. For controls we matched to: age, poverty income ratio > 1 (yes/no), smoker status (smokes now/does not), insurance status (has/does not), diabetes status (positive/negative), age of menopause, survey year ≥ 2005 (binary), black ethnicity (yes/no) and education ≥ “some college” (yes/no). Black ethnicity was included as black females had much higher BMD than non-black (U-test $p=10^{-16}$, mean difference: 0.75±0.02 z-score total BMD). Score balance was determined both visually using absolute mean difference, and KS threshold of 0.1.[41] All covariates were balanced in the after menopause control group, but age differed in the before menopause control. Pre-menopause control group differed by their median (IQR) of 51 years (48-53 years) versus 61 years (54-67 years) for the HRT group; this was expected and tolerated since the HRT are post-menopausal and therefore typically older.

**Regression fit lines** The regression lines in Figs 1, 2, 4 and 5 were computed using model averages weighted by repeated cross validation. To do so we performed model selection (models listed below) using 10-fold, 10-repeat cross-validation. We used the out-of-sample mean-squared error to choose continuous models, and AUC[42] (area under the ROC curve) to choose binomial models. Each fold included fitting the model to each of the FMP age estimates from YoGlen then pooling using Rubin's rules[38]; pooled individual predictions were used to compute the error. In Figs 1 and 2 we present the model average where we picked one of 4 Gaussian regression models at each repeat (10-knot spline, 10-knot spline with jump, 10-knot location-scale, or piecewise 10-knot location-scale) then aggregated grid predictions from -25 to 25 years-to-FMP over the 100 total repeats. The location-scale models included joint prediction of mean and variance to mitigate issues with heteroskedasticity, likely originating from late menopause transition fluctuations[21] (e.g., Fig 4 g-i). Similarly, in Fig 5 we present the model average where we picked one of 2 weighted binomial regression models at each repeat because the outcome variables are binary and location-scale models are therefore not relevant (10-knot spline, 10-knot spline with jump). Rare outcomes risk underfitting and therefore we set binomial case weights to 1 for the majority outcome and majority / minority ratio for minority outcome[43]. In Fig 4 we present data for females with known non-menopausal status, and therefore did not require model selection. We present the 10-knot Gaussian location-scale model which allows estimates of STD. Best fit lines used the `mgcv` package in R[44]. In analysis of HRT (Fig 3) we used linear regression on data pooled using Rubin's rules due to the relatively small number of data points.

### Statistics

Statistics were performed in R v4.5.1 (NHANES) and Python v3.9.23 (Clalit). All error bars/bands are standard errors. Correlations are Pearson's. All tests are two-sided. Significance was set at $p<0.05$. False discovery rate (FDR) correction used the Benjamini-Hochberg algorithm[45].

All methods are available on GitHub: https://github.com/AlonLabWIS/YoGlen.

---

**Tables and full resolution figures** are available on FigShare: https://doi.org/10.6084/m9.figshare.30597011.

### Data Availability

Aggregated and simulated data are available on GitHub: https://github.com/AlonLabWIS/YoGlen. Individualized data for NHANES can be downloaded from https://wwwn.cdc.gov/nchs/nhanes/. Individualized data from Clalit are not publicly available. Quantiles for the healthy, no meds Clalit data are available at https://tanaylab.weizmann.ac.il/labs/.

### Code Availability

All analysis code is available on Github: https://github.com/AlonLabWIS/YoGlen.

**Acknowledgements**

We thank all members of our labs and Gideon Kopernik, Keren Doenyas-Barak and Jaron Rabinovici for discussions. Clalit data usage was approved by the Clalit Helsinki Committee RMC-1059-20.

**Funding Statement**

This work was supported by the European Research Council (ERC) under the European Union’s Horizon 2020 research and innovation program (Grant Agreement No 856487), the Sagol Institute for Longevity Research and the The Knell Family Institute for Artificial Intelligence at the Weizmann Institute of Science. This work was supported in part by the Zuckerman STEM Leadership Program (GP). UA is the incumbent of the Abisch-Frenkel Professional Chair.

**Author contributions:**

All authors reviewed the manuscript prior to submission.
Conceptualization: UA, GP, NLS, MN, NH, YT
Methodology: UA, GP, YH, NH, YT
Formal Analysis: YH, GP
Funding acquisition: UA
Visualization: YH, GP, UA
Supervision: UA, NLS
Writing: UA, GP, YH

**Competing interests**

The authors declare that they have no competing interests.

| | Median (IQR) or prevalence (SE) | N obs | Missing | Median (IQR) or prevalence (SE) | N obs | Missing |
|---|---|---|---|---|---|---|
| **NHANES** | **Female** | | | **Male** | | |
| Age (years) | 51 (36-66) | 36287 | 0.0% | 51 (35-66) | 34440 | 0.0% |
| BMI ($kg/m^2$) | 28 (24-34) | 32896 | 9.3% | 28 (24-33) | 31398 | 8.8% |
| Smoke | 16% (0%) | 36235 | 0.1% | 24% (0%) | 34386 | 0.2% |
| Post menopause | 48% (0%) | 26567 | 14.0% | - | - | - |
| Menopause age | 50 (45-53) | 11412 | 20.0%[†] | - | - | - |
| Current HRT use* | 9% (0%) | 11443 | 38.6% | - | - | - |
| Current birth control use** | 6% (0%) | 3783 | 72.7% | - | - | - |
| **Clalit** | **Female** | | | **Male** | | |
| Age (years) | 49 (35-62) | 1.32 M | 0.0% | 55 (41-65) | 1.2 M | 0.0% |
| BMI ($kg/m^2$) | 28 (24-33) | 923 K | 30% | 28 (25-31) | 886 K | 26% |

**Table 1. Demographic summary.** [†] self-reported post-menopausal females only. * self-reported menopausal females only; varies greatly by study year, see Supplementary Fig S6. ** self-reported non-menopausal females only; self-reported birth control use or reported use of estrogen/progestin contraceptive.

## Supplementary information

Note: tables and full resolution figures are available on FigShare at https://figshare.com/account/articles/30597011.

**Table S1. Lab tests for NHANES (females).**

*$y = (log(x + \epsilon) - position)/scale$ where y is the preprocessed lab value and $x$ is the raw value (position and scale are the position and STD of $log(x + \epsilon)$ over ages 25-45; $\epsilon$ is to avoid log(0), it is equal to the smallest non-zero value when zero values are present). †|y| > 5. Note that the name LBXHCT was recycled (from hematocrit to hydroxycotinine) for waves 2013-2014, 2015-2016 and 2017-2018, that were excluded from analysis.

| Lab test | Short name | System | N obs | N outliers dropped† | Position* | Scale* | $\epsilon$* | NHANES code | Units |
|---|---|---|---|---|---|---|---|---|---|
| 17α-hydroxyprogesterone | 17-ohp | gonad | 3009 | 0 | 0.389 | 1.052 | 0 | LBD17HSI | nmol/L |
| albumin | albumin | liver | 29439 | 13 | 3.706 | 0.102 | 0 | LBDSALSI | g/L |
| alkaline phosphatase | alp | liver | 26978 | 13 | 4.139 | 0.333 | 0 | LBXSAPSI | U/L |
| alanine aminotransferase | alt | liver | 29395 | 42 | 2.875 | 0.464 | 0 | LBXSATSI | U/L |
| anti-mullerian hormone | amh | gonad | 3054 | 0 | 2.092 | 1.373 | 0 | LBDAMHSI | pmol/L |
| androstenedione | andro | gonad | 3053 | 1 | 1.209 | 0.496 | 0 | LBDANDSI | nmol/L |
| apolipoprotein | apob | lipid metabolism | 7621 | 2 | -0.168 | 0.289 | 0 | LBDAPBSI | g/L |
| aspartate aminotransferase | ast | liver | 29375 | 107 | 3.003 | 0.330 | 0 | LBXSASSI | U/L |
| vitamin b12 | b12 | micronutrients | 14530 | 34 | 5.857 | 0.510 | 0 | LBDB12SI | pmol/L |
| beta-2 microglobulin | b2m | kidney | 6451 | 20 | 0.523 | 0.219 | 0 | SSB2M | mg/L |
| bicarbonate | bicarb | kidney | 29401 | 9 | 3.165 | 0.095 | 0 | LBXSC3SI | mmol/L |
| bilirubin (total) | bilirubin | liver | 26966 | 21 | -0.724 | 0.517 | 0.01 | LBXSTB | mg/dL |
| body mass index | bmi | glucose metabolism | 34280 | 0 | 3.354 | 0.253 | 0 | BMXBMI | kg/m^2 |
| nt-probnp | bnp | heart | 6423 | 2 | 3.759 | 0.837 | 0 | SSBNP | pg/ml |
| bone alkaline phosphatase | bone alp | bone | 3374 | 0 | 2.539 | 0.391 | 0 | LBXBAP | ug/L |
| blood urea nitrogen | bun | kidney | 29431 | 6 | 2.302 | 0.355 | 0 | LBXSBU | mg/dL |
| calcium (total) | calcium | bone | 29422 | 18 | 0.838 | 0.039 | 0 | LBDSCASI | mmol/L |
| creatine phosphokinase | ck | muscle | 14538 | 18 | 4.530 | 0.558 | 0 | LBXSCK | IU/L |
| cobalt | cobalt | metal | 6459 | 21 | 1.217 | 0.637 | 0 | LBDBCOSI | nmol/L |
| c-peptide | c-peptide | glucose metabolism | 3386 | 4 | -0.378 | 0.478 | 0 | LBXCPSI | nmol/L |
| c-reactive protein | crp | inflammation | 15019 | 0 | -1.443 | 1.354 | 0 | LBXCRP | mg/dL |
| cystatin c | cystatin | kidney | 6455 | 23 | -0.420 | 0.199 | 0 | SSCYST | mg/L |
| dehydroepiandrosterone sulfate | dhea | gonad | 3030 | 2 | 1.010 | 0.631 | 0 | LBDDHESI | µmol/L |
| estrone | e1 | gonad | 3041 | 6 | 5.480 | 0.830 | 0 | LBDESOSI | pmol/L |
| estradiol | e2 | gonad | 8298 | 0 | 4.188 | 1.247 | 0 | LBXEST | pg/mL |
| eosinophils number | eosinophils | wbc | 33115 | 22 | -1.373 | 0.435 | 0.1 | LBDEONO | count |
| glucose (plasma, fasting) | f glucose | glucose metabolism | 16343 | 104 | 4.565 | 0.183 | 0 | LBXGLU | mg/dL |
| femoral neck bone mineral density | femor neck bmd | bone mass | 10767 | 2 | -0.157 | 0.150 | 0 | DXXNKBMD | gm/cm^2 |
| ferritin | ferritin | iron | 19452 | 0 | 3.570 | 0.982 | 0 | LBDFERSI | ug/L |
| fibrinogen | fibrinogen | clotting | 2862 | 2 | 1.259 | 0.214 | 0 | LBDFBSI | g/L |
| folate (serum, total) | folate | micronutrients | 15918 | 3 | 2.690 | 0.546 | 0 | LBDFOT | ng/mL |
| folate (rbc) | folate (rbc) | micronutrients | 16067 | 1 | 6.113 | 0.396 | 0 | LBDRFO | ng/mL |
| follicle stimulating hormone | fsh | gonad | 4826 | 0 | 1.685 | 1.070 | 0 | LBXFSH | mIU/mL |
| triiodothyronine (T3, free) | ft3 | thyroid | 4447 | 11 | 1.145 | 0.149 | 0 | LBXT3F | pg/mL |
| thyroxine (T4, free) | ft4 | thyroid | 4449 | 12 | 2.291 | 0.199 | 0 | LBDT4FSI | pmol/L |
| gamma glutamyl transferase | ggt | liver | 29434 | 38 | 2.819 | 0.601 | 0 | LBXSGTSI | U/L |
| glycohemoglobin | hba1c | glucose metabolism | 33179 | 188 | 1.676 | 0.119 | 0 | LBXGH | % |

| glucose tolerance test (2 hour) | gtt | glucose metabolism | 5482 | 0 | 1.751 | 0.324 | 0 | LBDGLTSI | mmol/L |
|---|---|---|---|---|---|---|---|---|---|
| high-density lipoprotein (direct) | hdl | lipid metabolism | 27902 | 3 | 0.348 | 0.276 | 0 | LBDHDDSI | mmol/L |
| hematocrit | hematocrit | rbc | 25111 | 34 | 3.651 | 0.093 | 0 | LBXHCT | % |
| hemoglobin | hemoglobin | iron | 33239 | 71 | 2.566 | 0.105 | 0 | LBXHGB | g/dL |
| homocysteine | homocysteine | kidney | 6903 | 3 | 1.782 | 0.339 | 0 | LBXHCY | umol/L |
| high sensitivity c-reactive protein | hscrp | inflammation | 12347 | 0 | 0.766 | 1.296 | 0 | LBXHSCRP | mg/L |
| insulin | insulin | glucose metabolism | 15945 | 7 | 2.270 | 0.721 | 0 | LBXIN | uU/mL |
| iron | iron | iron | 29416 | 3 | 2.489 | 0.542 | 0 | LBDSIRSI | umol/L |
| l1 bone mineral density | l1 bmd | bone mass | 10901 | 2 | -0.036 | 0.138 | 0 | DXXL1BMD | gm/cm^2 |
| l2 bone mineral density | l2 bmd | bone mass | 10502 | 1 | 0.065 | 0.126 | 0 | DXXL2BMD | gm/cm^2 |
| l3 bone mineral density | l3 bmd | bone mass | 10537 | 0 | 0.095 | 0.121 | 0 | DXXL3BMD | gm/cm^2 |
| l4 bone mineral density | l4 bmd | bone mass | 9965 | 1 | 0.083 | 0.120 | 0 | DXXL4BMD | gm/cm^2 |
| lactate dehydrogenase | ldh | muscle | 26852 | 20 | 4.834 | 0.210 | 0 | LBXSLDSI | U/L |
| low-density lipoprotein | ldl | lipid metabolism | 13928 | 4 | 1.001 | 0.305 | 0 | LBDLDLSI | mmol/L |
| lead | lead | metal | 30427 | 3 | -3.234 | 0.623 | 0.002 | LBDBPBSI | umol/L |
| luteinizing hormone | lh | gonad | 3054 | 0 | 1.697 | 1.185 | 0 | LBXLUH | mIU/mL |
| lymphocyte number | lymphocytes | wbc | 33115 | 17 | 0.763 | 0.305 | 0 | LBDLYMNO | count |

| mean arterial pressure* | map | heart | 25626 | 3 | 4.421 | 0.125 | 0 | bpd-pp/3 | mmHg |
|---|---|---|---|---|---|---|---|---|---|
| mean cell volume | mcv | rbc | 33239 | 98 | 4.471 | 0.079 | 0 | LBXMCVSI | fL |
| monocyte number | monocytes | wbc | 33115 | 11 | -0.505 | 0.276 | 0.1 | LBDMONO | count |
| segmented neutrophils number | neutrophils | wbc | 33115 | 9 | 1.449 | 0.405 | 0 | LBDNENO | count |
| serum neurofilament light chain | nfl | cognition | 1081 | 1 | 2.248 | 0.584 | 0 | SSSNFL | pg/ml |
| osmolality | osmolality | kidney | 29431 | 20 | 5.620 | 0.017 | 0 | LBXSOSSI | mOsm/kg |
| oxychlordane | oxychlordane | pop | 2063 | 0 | -3.018 | 0.713 | 0 | LBXOXY | ng/g |
| pcb074 | pcb074 | pop | 2169 | 0 | -3.494 | 0.628 | 0 | LBX074 | ng/g |
| phosphorus | phosphate | bone | 29431 | 3 | 0.170 | 0.151 | 0 | LBDSPHSI | mmol/L |
| platelet count | platelets | clotting | 33238 | 42 | 5.585 | 0.257 | 0 | LBXPLTSI | % |
| potassium | potassium | kidney | 29431 | 9 | 1.359 | 0.076 | 0 | LBXSKSI | mmol/L |
| pulse pressure | pp | heart | 25626 | 1 | 3.747 | 0.249 | 0 | sbp-dbp | mmHg |
| progesterone | progesterone | gonad | 3027 | 0 | 0.190 | 2.483 | 0 | LBDPG4SI | nmol/L |
| parathyroid hormone (Elecys method) | pth | bone | 4626 | 0 | 3.566 | 0.497 | 0 | LBXPT21 | pg/mL |
| red blood cell count | rbc | rbc | 33239 | 15 | 1.484 | 0.089 | 0 | LBXRBCSI | count |
| red blood cell distribution width | rdw | rbc | 33239 | 90 | 2.588 | 0.110 | 0 | LBXRDW | % |
| systolic blood pressure | sbp | heart | 25767 | 0 | 4.720 | 0.115 | 0 | BPXSY | mmHg |
| creatinine (serum) | scr | kidney | 29435 | 88 | 4.095 | 0.239 | 0 | LBDSCRSI | umol/L |
| selenium (blood) | selenium | metal | 15255 | 26 | 5.216 | 0.130 | 0 | LBXBSE | ug/L |
| sex hormone binding globulin | shbg | liver | 8083 | 0 | 4.109 | 0.695 | 0 | LBXSHBG | nmol/L |

| sodium | sodium | kidney | 29439 | 26 | 4.932 | 0.016 | 0 | LBXSNASI | mmol/L |
|---|---|---|---|---|---|---|---|---|---|
| testosterone (total) | testo | gonad | 10903 | 18 | 3.154 | 0.562 | 0 | LBXTST | ng/dL |
| thyroglobulin | thyroglobul in | thyroid | 4445 | 0 | 2.314 | 1.121 | 0 | LBDTGNSI | ug/L |
| total iron binding capacity | tibc | iron | 14416 | 3 | 4.180 | 0.189 | 0 | LBDTIBSI | umol/L |
| total bone mineral content | total bmc | bone mass | 5481 | 0 | 7.653 | 0.153 | 0 | DXDTOBM C | g |
| total bone mineral density | total bmd | bone mass | 5481 | 0 | 0.089 | 0.085 | 0 | DXDTOBM D | g/cm^2 |
| cholesterol (total) | total chol | lipid metabolism | 32547 | 5 | 5.225 | 0.201 | 0 | LBXTC | mg/dL |
| transferrin saturation | transferrin sat. | iron | 12213 | 5 | 2.933 | 0.607 | 0 | LBDPCT | % |
| triglycerides | tri | lipid metabolism | 29419 | 4 | 0.155 | 0.582 | 0 | LBDSTRSI | mmol/L |
| cardiac hs-troponin I (Abbott) | troponin | heart | 6435 | 1 | 0.074 | 0.753 | 0.1 | SSTNIA | ng/L |
| thyroid stimulating hormone | tsh | thyroid | 4448 | 18 | 0.329 | 0.837 | 0 | LBXTSH1 | uIU/mL |
| triiodothyronin e (T3, total) | tt3 | thyroid | 4448 | 2 | 0.574 | 0.294 | 0 | LBDTT3SI | nmol/L |
| thyroxine (T4, total) | tt4 | thyroid | 4441 | 5 | 2.071 | 0.206 | 0 | LBXTT4 | ug/dL |
| albumin (urine) | u albumin | kidney | 30747 | 31 | 2.094 | 1.212 | 0 | URXUMA | ug/mL |
| creatinine (urine) | u creatinine | kidney | 30721 | 0 | 4.604 | 0.732 | 3 | URXUCR | mg/dL |
| white blood cell count | wbc | wbc | 33238 | 10 | 1.987 | 0.296 | 0 | LBXWBCS I | count |

**Table S2. Lab tests for Clalit (females).**
*$y = (log(x) - position)/scale$ where y is the preprocessed lab value and $x$ is the raw value (position and scale are the position and STD of log(x) over ages 25-45). †|y| > 5. Values $x \leq 0$ were also excluded from analysis.

| Lab test | Short name | System | N obs | N outliers dropped[†] | Position* | Scale* | Units |
|---|---|---|---|---|---|---|---|
| fibrinogen | fibrinogen | clotting | 116293 | 35 | 5.849 | 0.274 | mg/dL |
| ferritin | ferritin | iron | 1719306 | 703 | 3.147 | 0.968 | ng/mL |
| iron | iron | iron | 2449977 | 242 | 4.176 | 0.581 | microg/dL |
| vitamin b12 | b12 | micronutrients | 3023218 | 52 | 5.616 | 0.372 | PMOL/L |
| folate (serum, total) | folate | micronutrients | 1519186 | 121 | 2.833 | 0.495 | nmol/L |
| albumin | albumin | liver | 4172796 | 15308 | 1.44 | 0.081 | g/dL |
| alanine aminotransferase | alt | liver | 6685744 | 17790 | 2.728 | 0.517 | units/L |
| aspartate aminotransferase | ast | liver | 6682873 | 28430 | 2.929 | 0.342 | units/L |
| bilirubin (total) | bilirubin | liver | 4497469 | 6260 | -0.695 | 0.457 | mg/dL |
| gamma glutamyl transferase | ggt | liver | 3118561 | 10286 | 2.816 | 0.606 | IU/L |
| cholesterol (total) | total chol | lipid metabolism | 6404325 | 1074 | 5.168 | 0.195 | mg/dL |
| high-density lipoprotein (direct) | hdl | lipid metabolism | 6028910 | 775 | 3.959 | 0.238 | mg/dL |
| low-density lipoprotein | ldl | lipid metabolism | 5934205 | 2006 | 4.612 | 0.282 | mg/dL |
| glucose (plasma, fasting) | f glucose | glucose metabolism | 7024278 | 31949 | 4.469 | 0.132 | mg/dL |
| glycohemoglobin | hba1c | glucose metabolism | 1101226 | 6581 | 1.673 | 0.087 | percent |
| triglycerides | tri | lipid metabolism | 6300450 | 1239 | 4.506 | 0.485 | mg/dL |
| body mass index | bmi | glucose metabolism | 1788096 | 735 | 3.257 | 0.24 | kg/m2 |
| insulin | insulin | glucose metabolism | 16250 | 3 | 2.243 | 0.759 | UIU/ML |

| creatine phosphokinase | ck | muscle | 2067366 | 4576 | 4.274 | 0.534 | IU/L |
|---|---|---|---|---|---|---|---|
| lactate dehydrogenase | ldh | muscle | 3199192 | 5064 | 5.741 | 0.246 | units/L |
| alkaline phosphatase | alp | liver | 5732857 | 8959 | 4.16 | 0.32 | units/L |
| phosphorus | phosphate | bone | 3704081 | 4065 | 1.27 | 0.153 | mg/dL |
| parathyroid hormone (Elecys method) | pth | bone | 97325 | 49 | 3.7459 | 0.627 | PG/ML |
| bicarbonate | bicarb | kidney | 43232 | 99 | 3.191 | 0.147 | mEq/L |
| sodium | sodium | kidney | 6086346 | 3803 | 4.938 | 0.016 | mEq/L |
| blood urea nitrogen | bun | kidney | 6657666 | 4091 | 3.162 | 0.28 | mg/dL |
| calcium (total) | calcium | bone | 4591709 | 5222 | 2.226 | 0.046 | mg/dL |
| creatinine (serum) | scr | kidney | 7138400 | 7003 | -0.418 | 0.187 | mg/dL |
| creatinine (urine) | u creatinine | kidney | 519270 | 116 | 4.743 | 0.667 | mg/dL |
| potassium | potassium | kidney | 328420 | 149 | 1.456 | 0.101 | mEq/L |
| hematocrit | hematocrit | rbc | 8299047 | 12623 | 3.642 | 0.099 | percent |
| hemoglobin | hemoglobin | iron | 8293168 | 18007 | 2.528 | 0.11 | g/dL |
| mean cell volume | mcv | rbc | 8306574 | 4986 | 4.443 | 0.085 | pg/cell |
| red blood cell distribution width | rdw | rbc | 7290162 | 17257 | 2.617 | 0.109 | % |
| red blood cell count | rbc | rbc | 8299961 | 11704 | 1.502 | 0.094 | K-cells/Î¼L |
| thyroid stimulating hormone | tsh | thyroid | 4771339 | 22247 | 0.595 | 0.61 | mIU/l |
| triiodothyronine (T3, free) | ft3 | thyroid | 604217 | 4995 | 1.593 | 0.203 | pmol/L |

| thyroxine (T4, free) | ft4 | thyroid | 1310752 | 6955 | 2.659 | 0.19 | pmol/L |
|---|---|---|---|---|---|---|---|
| triiodothyronine (T3, total) | tt3 | thyroid | 142592 | 72 | 0.776 | 0.346 | NMOL/L |
| thyroglobulin | thyroglobulin | thyroid | 1562827 | 542 | 1.091 | 0.141 | g/dL |
| monocyte number | monocytes | wbc | 8066397 | 4923 | -0.972 | 0.379 | K-cells/Î¼L |
| white blood cell count | wbc | wbc | 8299109 | 4029 | 1.902 | 0.305 | K-cells/Î¼L |
| eosinophils number | eosinophils | wbc | 8124619 | 100179 | 0.773 | 0.736 | K-cells/Î¼L |
| lymphocyte number | lymphocytes | wbc | 8096815 | 9345 | 0.716 | 0.337 | K-cells/Î¼L |
| segmented neutrophils number | neutrophils | wbc | 8101286 | 3567 | 1.338 | 0.409 | 1000/microL |
| estradiol | e2 | gonad | 1193933 | 12 | 6.23 | 1.18 | pmol/L |
| follicle stimulating hormone | fsh | gonad | 697924 | 193 | 2.031 | 0.778 | IU/L |
| luteinizing hormone | lh | gonad | 1057952 | 44 | 1.749 | 0.941 | IU/L |
| progesterone | progesterone | gonad | 808090 | 434 | 1.418 | 1.272 | NMOL/L |
| androstenedione | andro | gonad | 70776 | 9 | 1.754 | 0.595 | NMOL/L |
| testosterone (total) | testo | gonad | 282095 | 389 | 0.169 | 0.517 | NMOL/L |
| sex hormone binding globulin | shbg | liver | 1562827 | 542 | 1.091 | 0.141 | g/dL |
| c-reactive protein | crp | inflammation | 1104572 | 10618 | -0.939 | 1.386 | mg/dL |
| 17-hydroxy progesterone | 17-ohp | gonad | 168142 | 193 | 0.946 | 0.735 | NMOL/L |

| dehydroepiandrosterone sulfate | dhea | gonad | 237437 | 28 | 1.314 | 0.57 | UMOL/L |
|---|---|---|---|---|---|---|---|
| osmolality | osmolality | kidney | 4903 | 20 | 5.66 | 0.043 | mOsm/kg |
| platelet count | platelets | clotting | 8290422 | 11552 | 5.555 | 0.27 | K/microL |

**Table S3. HRT usage in NHANES by variable**, minimum 10 observation pairs. Population totals for menopausal females with all needed variables, includes: combo pill (or both estrogen and progestin): 127, estrogen pill: 512, progestin pill: 9, unknown: 4, not on hrt: 7110. *Mann-Whitney U test (paired). We used paired tests for p values and effect sizes in this table (N pairs), whereas we used unpaired values for Fig 3 (N HRT and N control), hence Fig 3 effect sizes differ slightly.

| Variable | System | N pairs | N HRT | N control | Control - HRT | Standard error | p (U test*) | p (FDR adj.) |
|---|---|---|---|---|---|---|---|---|
| sodium | kidney | 673 | 708 | 702 | -0.371 | 0.061 | 3E-08 | 3E-07 |
| potassium | kidney | 674 | 709 | 702 | -0.153 | 0.065 | 0.02 | 0.04 |
| bicarb | kidney | 673 | 709 | 701 | -0.171 | 0.052 | 8E-04 | 0.002 |
| u creatinine | kidney | 720 | 732 | 725 | 0.067 | 0.053 | 0.3 | 0.4 |
| cystatin | kidney | 304 | 378 | 384 | -0.397 | 0.083 | 4E-06 | 3E-05 |
| b2m | kidney | 304 | 378 | 384 | -0.353 | 0.095 | 5E-04 | 0.002 |
| osmolality | kidney | 676 | 710 | 703 | -0.458 | 0.063 | 4E-12 | 1E-10 |
| scr | kidney | 671 | 708 | 700 | -0.022 | 0.056 | 1 | 1 |
| bun | kidney | 676 | 710 | 703 | -0.125 | 0.053 | 0.01 | 0.03 |
| u albumin | kidney | 719 | 731 | 725 | -0.076 | 0.054 | 0.09 | 0.2 |
| dhea | gonad | 0 | 0 | 0 | - | - | - | - |
| 17-ohp | gonad | 7 | 23 | 42 | - | - | - | - |
| andro | gonad | 7 | 23 | 42 | - | - | - | - |
| calcium | bone | 675 | 710 | 702 | -0.228 | 0.056 | 6E-05 | 3E-04 |
| phosphate | bone | 676 | 710 | 703 | -0.275 | 0.052 | 2E-07 | 2E-06 |
| pth | bone | 50 | 150 | 165 | -0.369 | 0.199 | 0.03 | 0.06 |
| alp | liver | 419 | 533 | 547 | -0.425 | 0.066 | 7E-10 | 1E-08 |
| bone alp | bone | 37 | 134 | 129 | -0.746 | 0.206 | 0.001 | 0.004 |
| nfl | cognition | 0 | 13 | 12 | - | - | - | - |
| testo | gonad | 33 | 87 | 121 | -0.841 | 0.307 | 0.02 | 0.04 |
| e2 | gonad | 31 | 67 | 116 | 0.628 | 0.209 | 0.003 | 0.009 |
| fsh | gonad | 56 | 159 | 174 | -0.458 | 0.159 | 9E-04 | 0.003 |
| lh | gonad | 7 | 23 | 42 | - | - | - | - |
| progsterone | gonad | 7 | 23 | 42 | - | - | - | - |
| shbg | liver | 31 | 67 | 118 | 0.256 | 0.205 | 0.2 | 0.3 |
| amh | gonad | 7 | 23 | 42 | - | - | - | - |
| e1 | gonad | 7 | 22 | 41 | - | - | - | - |
| troponin | heart | 306 | 380 | 383 | -0.291 | 0.074 | 1E-04 | 6E-04 |
| bnp | heart | 303 | 377 | 383 | 0.090 | 0.095 | 0.1 | 0.2 |
| ck | muscle | 70 | 132 | 186 | -0.160 | 0.169 | 0.4 | 0.5 |
| apob | lipid metabolism | 40 | 145 | 117 | -0.065 | 0.156 | 0.7 | 0.8 |
| homocysteine | kidney | 140 | 270 | 285 | -0.446 | 0.120 | 5E-04 | 0.002 |
| sbp | heart | 639 | 693 | 671 | -0.004 | 0.074 | 1 | 1 |
| map | heart | 635 | 691 | 668 | 0.006 | 0.057 | 0.9 | 1 |
| pp | heart | 635 | 691 | 668 | -0.001 | 0.075 | 0.8 | 0.9 |
| fibrinogen | clotting | 163 | 288 | 267 | -0.323 | 0.104 | 0.005 | 0.01 |
| crp | inflammation | 448 | 581 | 521 | 0.318 | 0.052 | 8E-10 | 1E-08 |
| hscrp | inflammation | 25 | 60 | 102 | 0.068 | 0.300 | 0.5 | 0.6 |
| neutrophils | wbc | 701 | 721 | 718 | 0.087 | 0.048 | 0.1 | 0.2 |
| monocytes | wbc | 702 | 721 | 719 | 0.046 | 0.052 | 0.4 | 0.6 |
| lymphocytes | wbc | 700 | 721 | 717 | 0.081 | 0.059 | 0.3 | 0.5 |
| platelets | clotting | 703 | 722 | 719 | 0.231 | 0.053 | 5E-05 | 3E-04 |
| wbc | wbc | 703 | 722 | 719 | 0.091 | 0.050 | 0.09 | 0.2 |
| eosinophils | wbc | 702 | 721 | 719 | -0.014 | 0.049 | 0.6 | 0.7 |
| alt | liver | 675 | 709 | 703 | -0.204 | 0.046 | 2E-06 | 2E-05 |
| ast | liver | 669 | 707 | 699 | -0.146 | 0.050 | 0.002 | 0.005 |
| ggt | liver | 673 | 710 | 700 | -0.204 | 0.059 | 9E-05 | 4E-04 |
| bilirubin | liver | 419 | 533 | 547 | -0.017 | 0.054 | 0.6 | 0.7 |
| albumin | liver | 676 | 710 | 703 | -0.176 | 0.039 | 2E-05 | 1E-04 |
| transferrin sat. | iron | 130 | 292 | 267 | 0.015 | 0.071 | 0.9 | 0.9 |
| iron | iron | 676 | 710 | 703 | 0.104 | 0.037 | 0.007 | 0.02 |
| ferritin | iron | 245 | 393 | 374 | -0.152 | 0.082 | 0.06 | 0.1 |
| b12 | micronutrients | 417 | 514 | 517 | -0.055 | 0.072 | 0.4 | 0.6 |
| folate | micronutrients | 48 | 113 | 157 | 0.040 | 0.225 | 0.9 | 1 |
| folate (rbc) | micronutrients | 51 | 114 | 161 | 0.015 | 0.217 | 0.8 | 0.9 |
| hemoglobin | iron | 705 | 722 | 721 | 0.019 | 0.040 | 0.7 | 0.8 |
| hematocrit | rbc | 567 | 655 | 618 | -0.016 | 0.051 | 0.6 | 0.7 |
| mcv | rbc | 704 | 722 | 720 | 0.132 | 0.036 | 0.001 | 0.004 |
| rdw | rbc | 703 | 722 | 719 | -0.246 | 0.036 | 2E-12 | 1E-10 |
| ldh | muscle | 418 | 532 | 547 | -0.414 | 0.066 | 2E-08 | 2E-07 |
| rbc | rbc | 705 | 722 | 721 | -0.156 | 0.054 | 0.006 | 0.01 |

| tibc | iron | 269 | 409 | 379 | 0.431 | 0.065 | 1E-09 | 1E-08 |
|---|---|---|---|---|---|---|---|---|
| f glucose | glucose metabolism | 164 | 368 | 351 | -0.172 | 0.115 | 0.03 | 0.05 |
| hba1c | glucose metabolism | 698 | 720 | 714 | -0.231 | 0.047 | 5E-07 | 4E-06 |
| total chol | lipid metabolism | 680 | 712 | 705 | -0.113 | 0.050 | 0.02 | 0.04 |
| ldl | lipid metabolism | 149 | 352 | 333 | -0.275 | 0.116 | 0.02 | 0.04 |
| hdl | lipid metabolism | 310 | 417 | 438 | 0.220 | 0.073 | 0.001 | 0.004 |
| tri | lipid metabolism | 676 | 710 | 703 | 0.021 | 0.047 | 0.6 | 0.7 |
| insulin | glucose metabolism | 160 | 364 | 343 | -0.253 | 0.102 | 0.03 | 0.06 |
| c-peptide | glucose metabolism | 82 | 201 | 197 | -0.307 | 0.158 | 0.04 | 0.08 |
| bmi | glucose metabolism | 707 | 724 | 719 | -0.180 | 0.044 | 4E-04 | 0.001 |
| gtt | glucose metabolism | 29 | 119 | 93 | -0.517 | 0.244 | 0.1 | 0.2 |
| tsh | thyroid | 21 | 105 | 69 | 0.184 | 0.222 | 0.7 | 0.8 |
| ft4 | thyroid | 21 | 105 | 69 | 0.152 | 0.264 | 0.6 | 0.7 |
| tt4 | thyroid | 21 | 105 | 69 | 0.436 | 0.267 | 0.1 | 0.2 |
| ft3 | thyroid | 21 | 105 | 69 | 0.283 | 0.307 | 0.4 | 0.6 |
| tt3 | thyroid | 21 | 105 | 69 | 0.130 | 0.321 | 0.6 | 0.7 |
| thyroglobulin | thyroid | 21 | 105 | 69 | -0.071 | 0.416 | 0.8 | 0.9 |
| selenium | metal | 48 | 119 | 149 | -0.260 | 0.173 | 0.2 | 0.3 |
| lead | metal | 647 | 705 | 675 | -0.420 | 0.048 | 7E-17 | 5E-15 |
| cobalt | metal | 27 | 63 | 107 | 0.034 | 0.187 | 0.9 | 1 |
| oxychlordane | pop | 23 | 110 | 128 | -0.130 | 0.195 | 0.4 | 0.6 |
| pcb074 | pop | 29 | 127 | 119 | -0.241 | 0.265 | 0.6 | 0.7 |
| total bmd | bone mass | 6 | 32 | 53 | - | - | - | - |
| total bmc | bone mass | 6 | 32 | 53 | - | - | - | - |
| femor neck bmd | bone mass | 111 | 235 | 191 | 0.301 | 0.151 | 0.04 | 0.08 |
| l1_bmd | bone mass | 108 | 227 | 190 | 0.604 | 0.150 | 4E-04 | 0.001 |
| l2_bmd | bone mass | 89 | 205 | 181 | 0.851 | 0.195 | 1E-04 | 5E-04 |
| l3_bmd | bone mass | 99 | 213 | 186 | 0.702 | 0.176 | 2E-04 | 7E-04 |
| l4_bmd | bone mass | 93 | 204 | 181 | 0.972 | 0.209 | 3E-05 | 2E-04 |

**Table S4. Clinical outcome variables in NHANES (females)**

| Variable | System | N obs | NHANES Code | description | notes |
|---|---|---|---|---|---|
| anemia | rbc | 36308 | MCQ053 | Taking treatment for anemia/past 3 mos | |
| angina | heart outcome | 36234 | MCQ160D | Ever told you had angina/angina pectoris | |
| arthritis | bone outcome | 36267 | MCQ160A | Doctor ever said you had arthritis | |
| asthma | lung | 36328 | MCQ010 | Ever been told you have asthma | |
| bronchitis | lung | 27322 | MCQ160K | Ever told you had chronic bronchitis | |
| copd | lung | 8691 | MCQ160O | Ever told you had COPD? | |
| depress score | depression | 23721 | DPQ0(1-9)0 | sum, NAs imputed with average | |
| diabetes | misc outcome | 36328 | DIQ010 | Doctor told you have diabetes | borderline or worse |
| diff falling asleep | sleep | 2486 | SLQ080 | How often have trouble falling asleep? | |
| emphysema | lung | 27345 | MCQ160G | Ever told you had emphysema | |
| ever smoke | smoking | 36306 | SMQ020 | Smoked at least 100 cigarettes in life | |
| gout | kidney outcome | 17604 | MCQ160N | Doctor ever told you that you had gout? | |
| heart attack | heart outcome | 36291 | MCQ160E | Ever told you had heart attack | |
| heart disease | heart outcome | 36212 | MCQ160C | Ever told you | |

| | | | | had coronary heart disease | |
|---|---|---|---|---|---|
| heart failure | heart outcome | 36266 | MCQ160B | Ever told had congestive heart failure | |
| insufficient sleep | sleep | 2158 | SLQ130 | How often did you not get enough sleep? | |
| kidney disease | kidney outcome | 36285 | KIQ02(0 and 2) | Ever told you had weak/failing kidneys | |
| kidney stones | kidney outcome | 22257 | KIQ026 | Ever had kidney stones? | |
| liver disease | misc outcome | 36286 | MCQ160L | Ever told you had any liver condition | |
| osteoporosis | bone outcome | 21878 | OSQ060 | Ever told had osteoporosis/brittle bones | |
| sleep breathing | sleep | 4861 | SLQ120 | How often feel overly sleepy during day? | |
| sleep disorder | sleep | 14154 | SLQ060 | Ever told by doctor have sleep disorder? | |
| sleeping pills | sleep | 4510 | SLQ140 | How often take pills to help you sleep? | |
| smoke now | smoking | 36304 | SMQ040 | Do you now smoke cigarettes | |
| snore | sleep | 8123 | SLQ030 | How often do you snore? | |
| stroke | heart outcome | 36303 | MCQ160F | Ever told you had a stroke | |
| thyroid problem | misc outcome | 31243 | MCQ160M | Ever told you had thyroid problem | |
| trouble sleeping | sleep | 24586 | SLQ050 | Ever told doctor had trouble sleeping? | |
| wake up at | sleep | 2279 | SLQ090 | How often wake | |

| night | | | | up during night? | |
|---|---|---|---|---|---|
| wake up early | sleep | 2211 | SLQ110 | How often feel unrested during the day? | |
| weak grip | functional | 4973 | MGXH(1-2)T(1-3) | Male cutoff: 29 kg<br>Female cutoff: 18 kg[46] | Max of 6 measurements = 3 trials x 2 hands |

**Table S5. HRT drug classes in NHANES**

We included prescription drug usage when considering HRT status. All drug classes labelled as e2 or combo were included as HRT (bolded). Note that we also included females who self-reported HRT usage without specifying the drug class.

| NHANES code (RXDDRGID) | Drug class (RXDDRUG) | Label |
|---|---|---|
| a10899 | ESTRADIOL; ESTRIOL; ESTRONE | e2 |
| a10900 | ESTRADIOL; ESTRIOL | e2 |
| a11518 | CONJUGATED ESTROGENS; PROGESTERONE | combo |
| a70248 | ESTRIOL | e2 |
| c00101 | SEX HORMONES - UNSPECIFIED | excluded |
| c00102 | CONTRACEPTIVES - UNSPECIFIED | contraceptive |
| c00183 | ESTROGENS - UNSPECIFIED | e2 |
| c00186 | SEX HORMONES COMBINATION - UNSPECIFIED | unknown (excluded) |
| d00229 | ETHINYL ESTRADIOL | e2 |
| d00534 | ESTRONE | e2 |
| d00537 | ESTRADIOL | e2 |
| d00541 | CONJUGATED ESTROGENS | e2 |
| d00542 | ESTERIFIED ESTROGENS | e2 |

| d00543 | ESTROPIPATE | e2 |
|---|---|---|
| d00550 | PROGESTERONE | progestin |
| d00554 | HYDROXYPROGESTERONE | progestin |
| d00555 | NORETHINDRONE | contraceptive |
| d00557 | LEVONORGESTREL | contraceptive |
| d03034 | NORGESTREL | progestin |
| d03238 | ETHINYL ESTRADIOL; NORETHINDRONE | combo |
| d03240 | MESTRANOL; NORETHINDRONE | combo |
| d03241 | ETHINYL ESTRADIOL; NORGESTREL | combo |
| d03242 | ETHINYL ESTRADIOL; LEVONORGESTREL | combo |
| d03245 | ESTERIFIED ESTROGENS; METHYLTESTOSTERONE | e2 |
| d03388 | ETHINYL ESTRADIOL; ETHYNODIOL | combo |
| d03389 | ESTRADIOL; TESTOSTERONE | e2 |
| d03781 | ETHINYL ESTRADIOL; NORGESTIMATE | combo |
| d03782 | DESOGESTREL; ETHINYL ESTRADIOL | combo |
| d03819 | CONJUGATED ESTROGENS; MEDROXYPROGESTERONE | combo |
| d04210 | ESTRADIOL TOPICAL | e2 |
| d04213 | PROGESTERONE TOPICAL | progestin |
| d04375 | ESTRADIOL; NORETHINDRONE | combo |
| d04396 | CONJUGATED ESTROGENS TOPICAL | e2 |
| d04506 | ESTRADIOL; NORGESTIMATE | combo |
| d04721 | ESTRADIOL; MEDROXYPROGESTERONE | combo |

| d04760 | DROSPIRENONE; ETHINYL ESTRADIOL | combo |
|---|---|---|
| d04772 | ETONOGESTREL | progestin |
| d04773 | ETHINYL ESTRADIOL; ETONOGESTREL | combo |
| d04779 | ETHINYL ESTRADIOL; NORELGESTROMIN | combo |
| d04914 | ESTRADIOL; LEVONORGESTREL | combo |
| d05027 | ESTRIOL | e2 |
| d05530 | DROSPIRENONE; ESTRADIOL | combo |
| d06272 | DIENOGEST; ESTRADIOL | combo |
| d07697 | DROSPIRENONE; ETHINYL ESTRADIOL; LEVOMEFOLATE | combo |

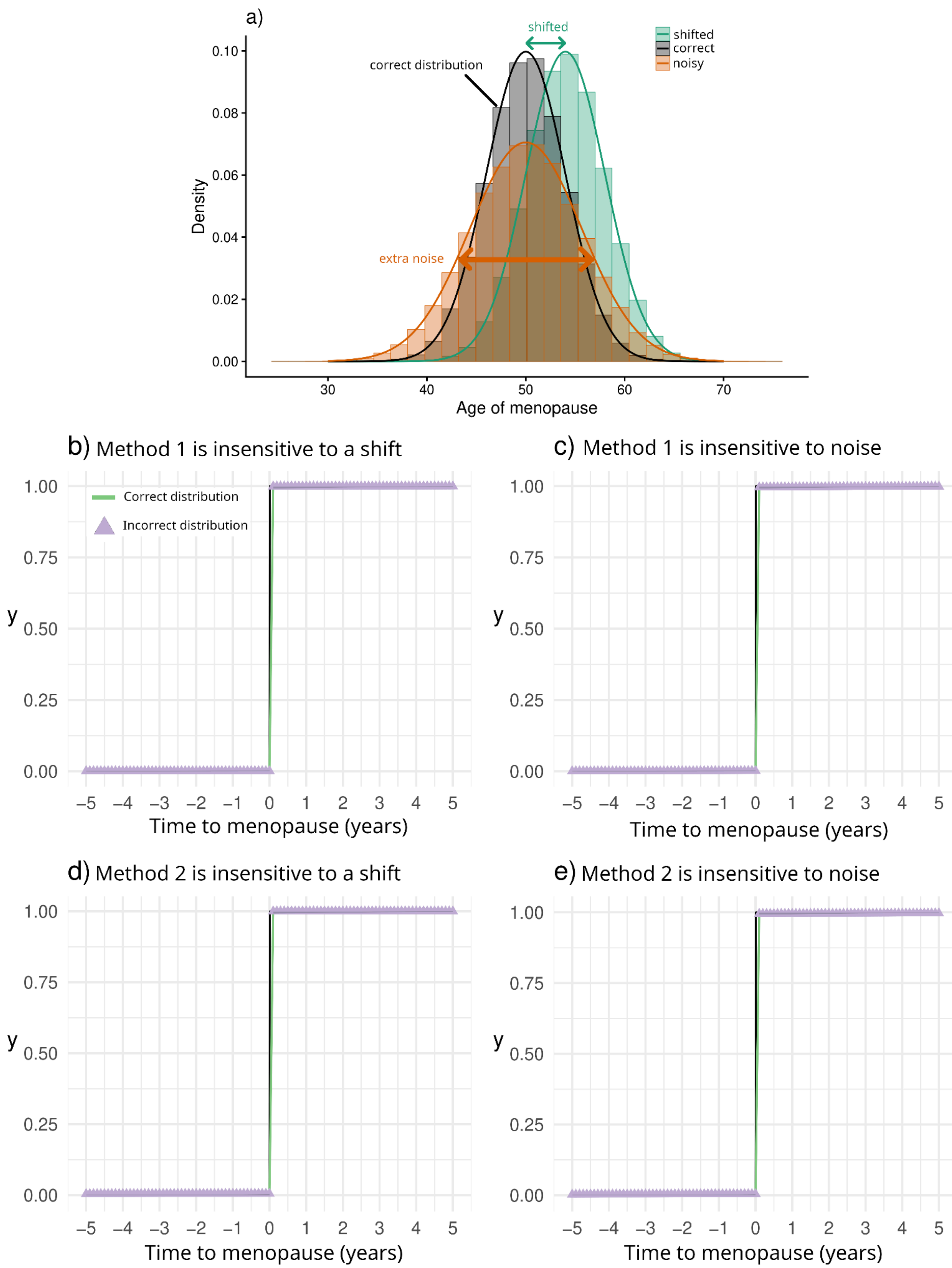

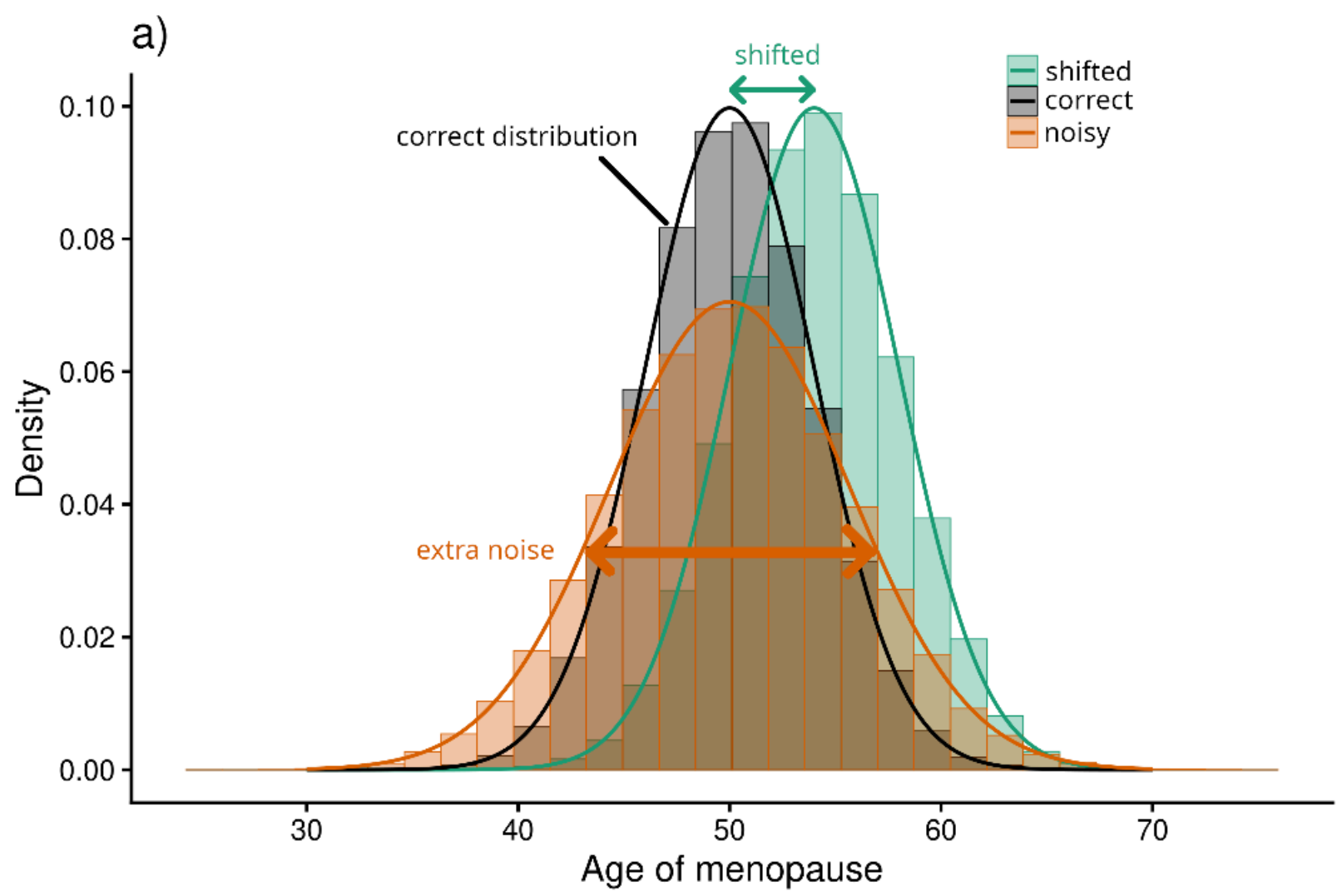

b) Method 1 is insensitive to a shift

Correct distribution

Incorrect distribution

y

Time to menopause (years)

c) Method 1 is insensitive to noise

y

Time to menopause (years)

d) Method 2 is insensitive to a shift

y

Time to menopause (years)

e) Method 2 is insensitive to noise

y

Time to menopause (years)

**Fig S1. The YoGlen method is insensitive to variations in the FMP distribution.** Simulated data was a step function at t=0 convolved with a normal FMP distribution $N(\mu_0, \sigma_0)$ with $\mu_0$=50 years and $\sigma_0$= 4 years. a) Original reference FMP distribution (black) and perturbed distributions: one that is shifted by 4 years (green, $\mu = \mu_0 + 4$) ( differences between mean FMP age in different countries span about 4 years[18]) and another that is wider due to added noise (orange, $\sigma^2 = \sigma_0^{\ 2} + 4^2$) (larger than typical FMP self-report error noise ~1 year[39,47]). b) deconvolving with YoGlen method 1 using an FMP distribution that has a shifted mean has negligible effect on YoGlen output. c) Deconvolving with YoGlen method 1 using an FMP distribution that is wider due to excess noise has negligible effect on YoGlen output. d) and e) same for YoGlen method 2.

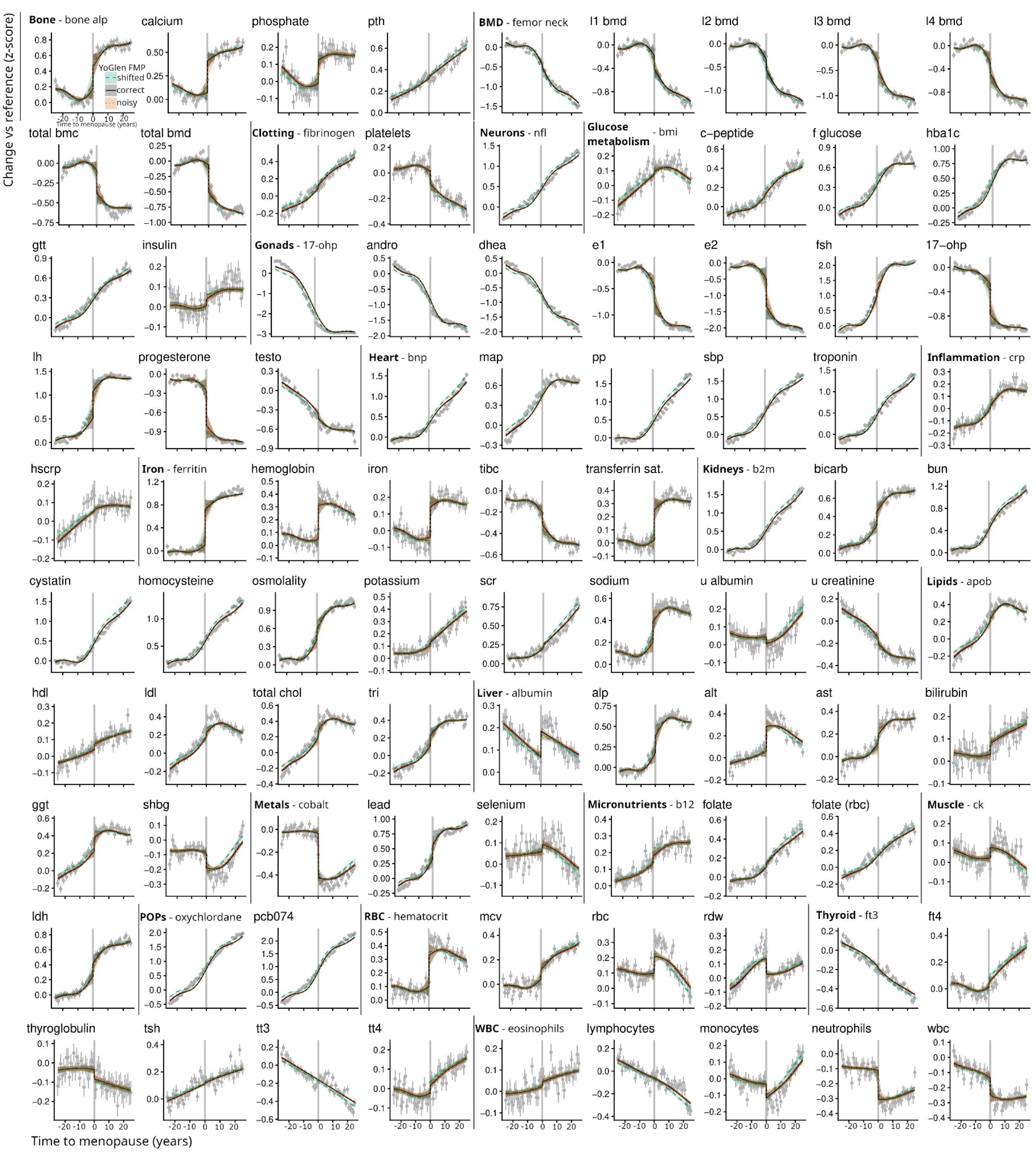

**Fig S2. Validation of YoGlen using simulated data.** We used YoGlen model 1 fits from Fig 2 for 90 tests as a reference to generate simulated data (gray points) by drawing age and FMP age from the NHANES distribution and generating lab test values according to the reference fits (correlations and variances were also matched to NHANES). The simulated population size was equal to the observed population size (N=36,356). We then hid the FMP ages larger than current age, and used YoGlen model 1 to reconstruct the fit as in Fig 2 (black, correct FMP distribution). The agreement in the unknown region (time to menopause < 0) with the reference fit was excellent, Pearson's r=0.97 ($p<10^{-15}$). To test sensitivity to errors in the FMP distribution,

we repeated the analysis using altered FMP distributions as in Fig S1a, with mean shifted later by 4 years (green curve), and excess noise (adding a random number to FMP from a normal distribution with mean 0 and STD of 4 years, orange curve). YoGlen was insensitive to these changes to FMP distribution, as evidenced by the excellent agreement with the original reference fit, Pearson's r=0.94 (shift FMP) and r=0.96 (noisy FMP) (both p<$10^{-15}$). This indicates YoGlen is robust against observational noise and errors in FMP age.

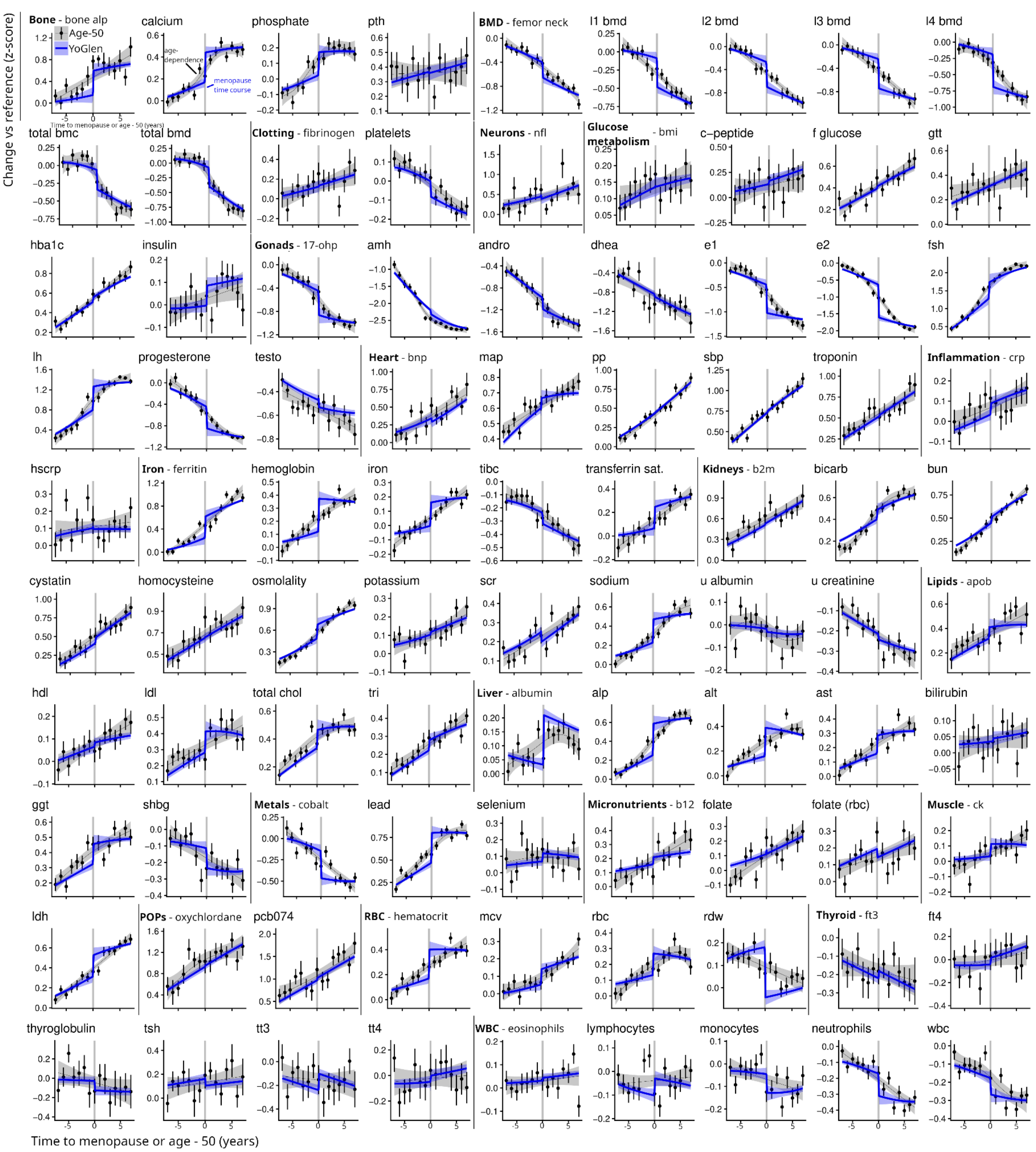

**Fig S3. Female lab tests show jumps at the FMP**, zoomed in to -7 to 7 years around the FMP (NHANES). Plotted are the cross-sectional mean test values per year of age versus age minus 50 (gray), and the YoGlen time-to-FMP where t=0 is the FMP (blue). Time=0 on the x axis corresponds to 50 years for the gray curves, and to the age of FMP for the blue curves. Black line is 10-knot spline fit to age; blue line is 10 fold, 10 repeat cross-validation model average fit to time to menopause, the same as in Fig 2.

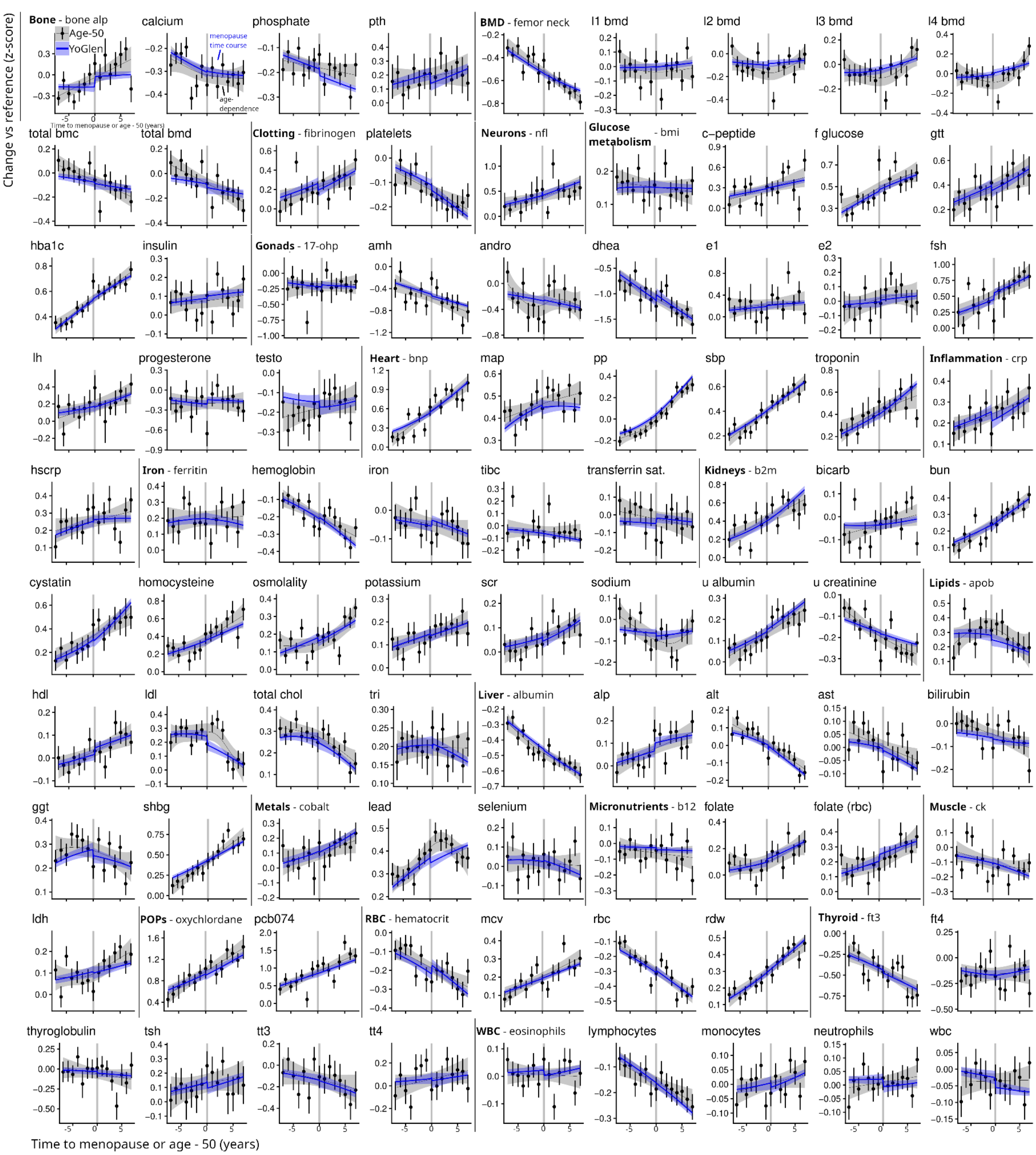

**Fig S4. Male lab tests do not show a jump** when analyzed with the same YoGlen algorithm and FMP distribution as females (NHANES). Plotted are the cross-sectional mean test values per year of age versus age minus 50 (gray), and the YoGlen time-to-FMP (blue). Black line is 10-knot spline fit to age; blue line is 10 fold, 10 repeat cross-validation model average fit to time to menopause, the same as in Fig 2.

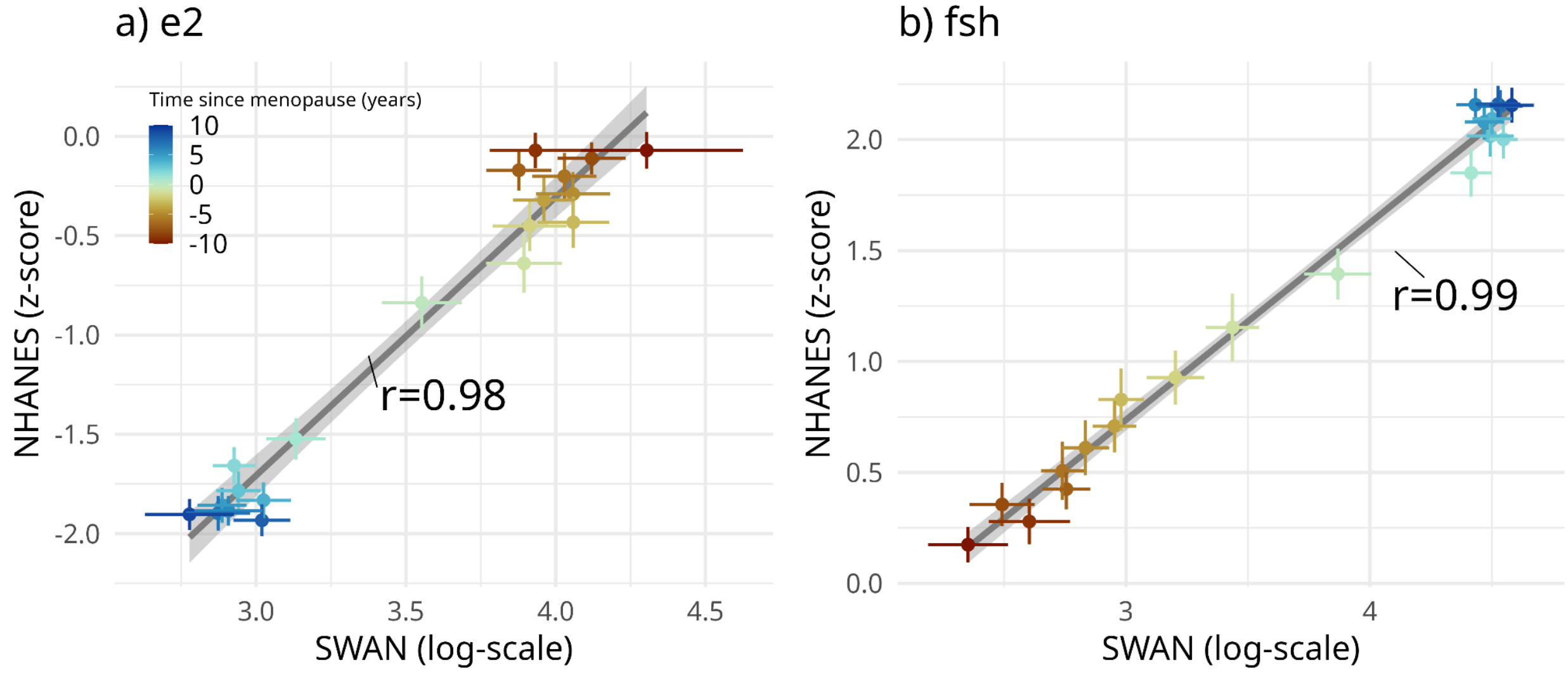

**Fig S5. Validation of YoGlen method (NHANES) by comparison to the longitudinal Study of Women's Health Across the Nation (SWAN) for the key sex hormones. a)** e2 (Pearson's r=0.98, $p=10^{-14}$.) **b)** fsh (r=0.99, $p<10^{-15}$). 1-year bins from -10 to 10 years before/after menopause. SWAN data are log-scaled; NHANES data are log-scaled then z-scored by healthy reference (Table S1).

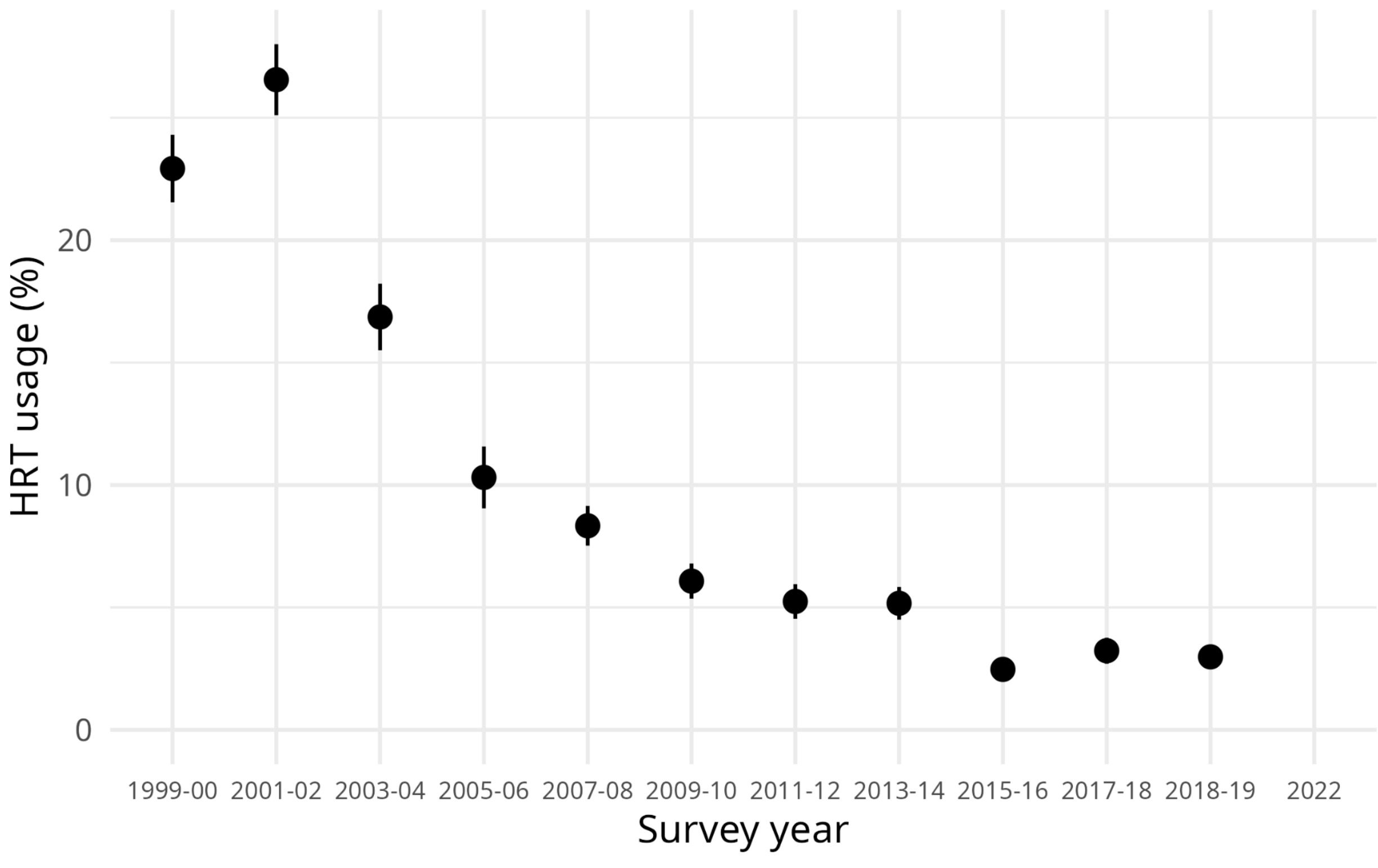

**Fig S6. HRT usage reported in NHANES dropped around 2004**, as seen globally in the wake of the WHI study[48]. HRT usage among post-menopausal females. Includes both self-reported

usage (asked until 2013-2014) and generic estradiol drug usage (asked each year except 2022). The overall frequency from 1999-2019 was 9.0%.

### Sensitivity analysis

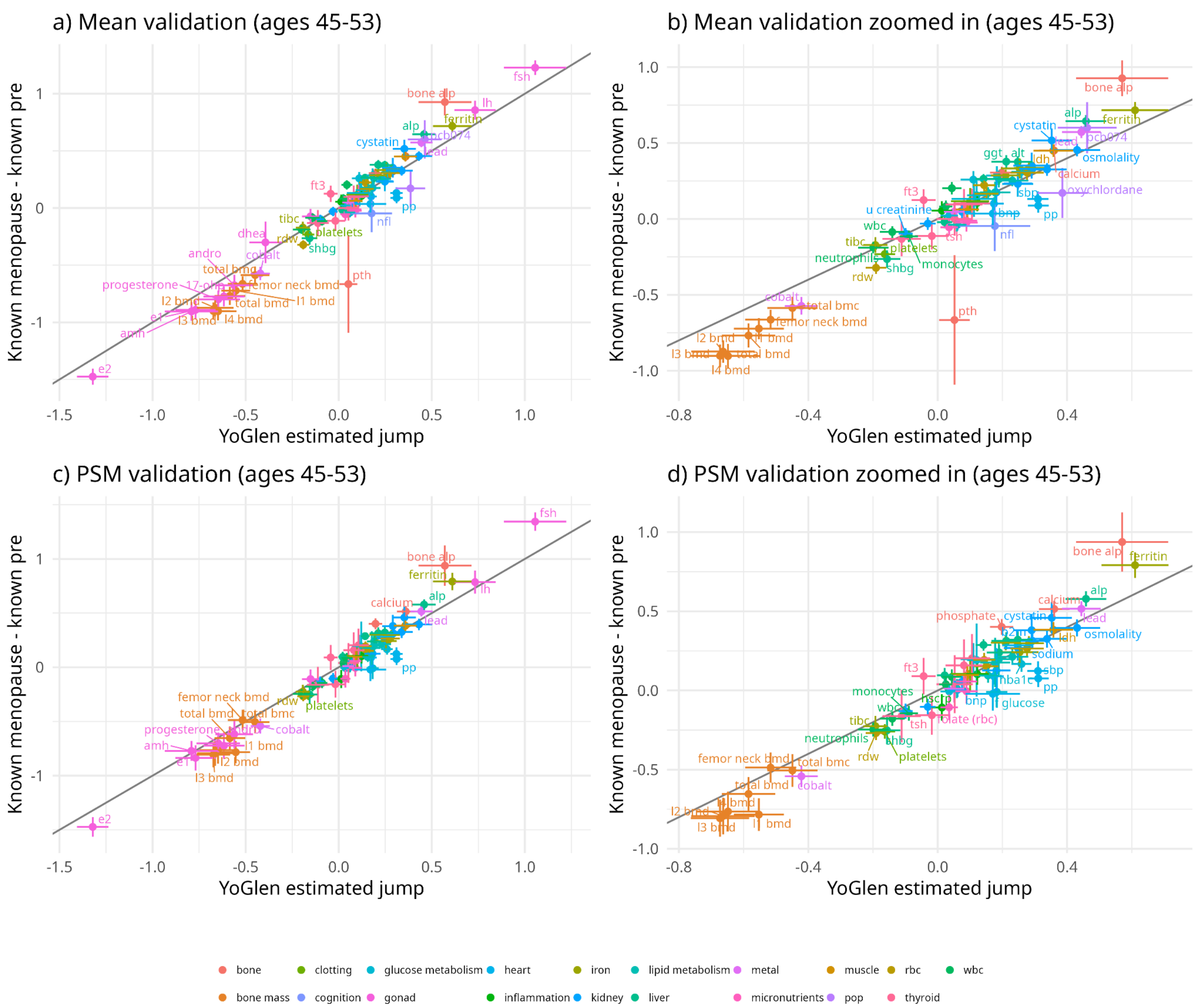

**Fig S7. Sensitivity analysis of jump sizes** shows strong concordance between YoGlen and direct estimates (NHANES). We compared the difference between pre-menopausal and post-menopausal females aged 45-53 who did not have a hysterectomy with known menopause status (y-axis) to the YoGlen estimates from Figure 2 (x-axis), both with (left) and without (right) the sex hormones. **a)** We see strong concordance between the mean difference and YoGlen estimate, with Pearon's rho = 0.97 (0.95-0.98 95% CI, $p=10^{-16}$; n=3155 individuals). **b)** Same as figure a) but zoomed in (i.e., excluding sex hormones). **c)** We see strong concordance with the mean PSM differences and the YoGlen estimate. Pearon's rho = 0.98 (0.97-0.99 95% CI, $p=10^{-16}$,n=890 pairs). **d)** Same as figure c) but zoomed in. Known menopause status only. PSM is paired difference after matching based on: poverty, smoking, insurance, diabetes, pre vs post 2005, black ethnicity and "some college" (or more) education level (all balanced, with absolute mean difference and KS threshold of 0.1[41]). Minimum 50 measurements per test.

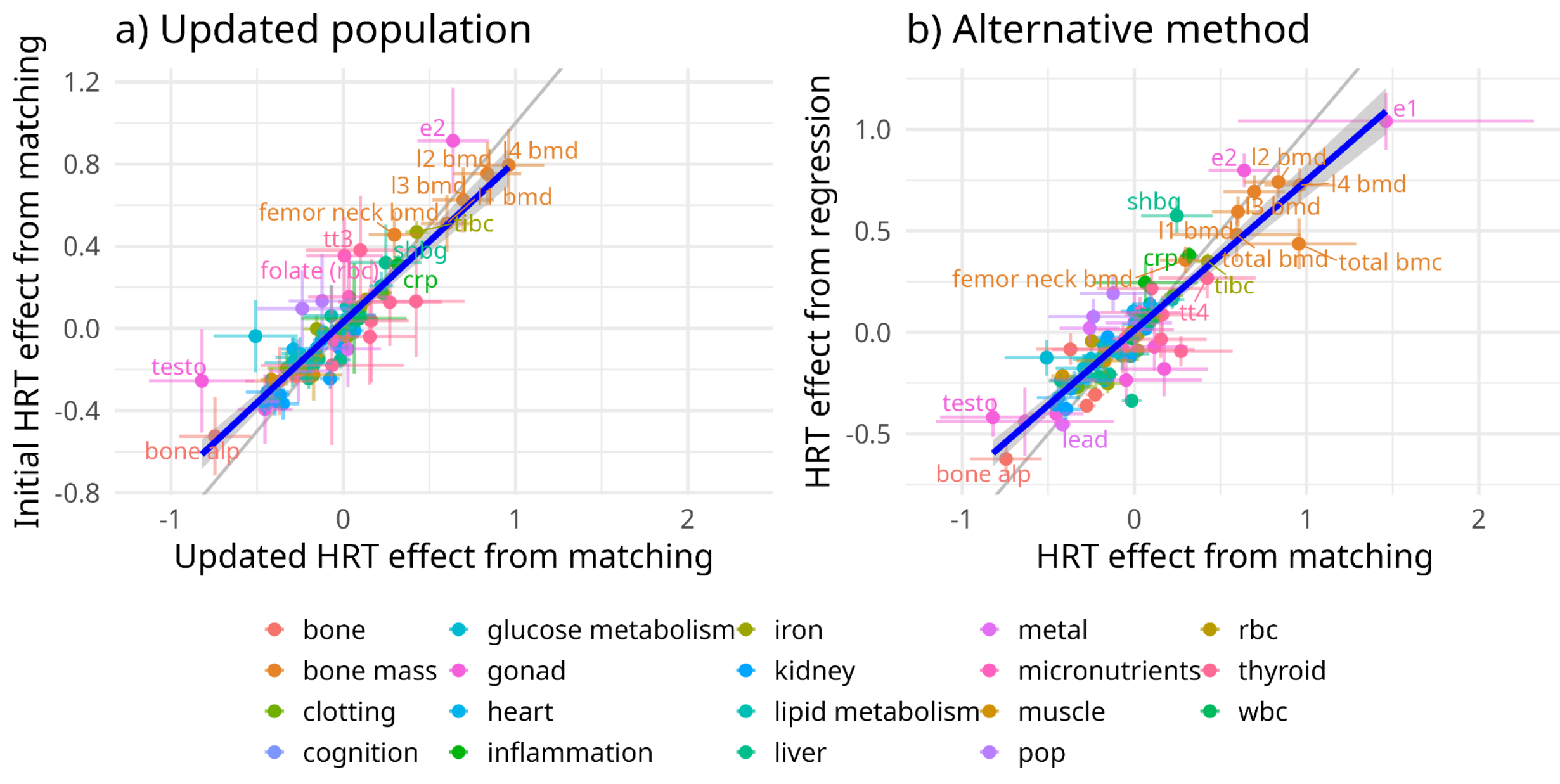

**Fig S8. Sensitivity analysis of HRT effect sizes** indicates that the results are robust to changing the population (a) and changing the estimating algorithm (b). **a)** Our initial effect sizes estimated using self-reported HRT status from years 1999-2012 only (y-axis) correlated strongly with our updated estimates including specific drugs used for years 1999-2019 (x-axis), r=0.90 (95% CI: 0.85-0.93, $p=10^{-16}$); propensity-score matched. **b)** Estimates using ordinary linear regression (y-axis) correlate strongly with our updated estimates r=0.90 (95% CI: 0.86-0.94, $p=10^{-16}$). Propensity-score matched using: survey period (pre vs post 2005), menopause status, age, poverty income ratio, smoker status, insurance status, diabetes status, black ethnicity (updated only) and education. Ordinary linear regression used: male age-dependence from 10-knot thin-plate spline[44], survey year, menopause status, age, poverty income ratio, smoker status, insurance status, diabetes status, all ethnicities and education.

## Supplementary discussion

### Why is it called a deconvolution and what is the relationship to latent variable modelling?

Our latent variable is the unknown age of menopause $a_m$, which we seek to infer from a proxy (age, $a$), so that we can understand its relationship with the observed lab value $y$. This is akin to the problem of deblurring an image where $a_m$ is the true image and $a$ is the distorted image observed through the lens. We don't know the latent variable so we have to invert what we do know which is

$$p(y|a) = \int_{-\infty}^{\infty} p_1(y|a_m, a) p_2(a_m|a) da_m.$$

In our case $y$ depends on the time to FMP

$$p(y|a) = \int_{-\infty}^{\infty} p_1(y|a - a_m)p_2(a_m|a)da_m = p_2 * p_1,$$

a convolution of functions $p_2$ and $p_1$ over the latent variable, $a_m$. We want to solve the inverse problem to find $p_1(y|a - a_m)$ based on the observed $p(y|a)$ and reference $p_2(a_m|a)$. Solving for $p_1(y|a - a_m)$ is thus a process of deconvolution, since it is the inverse of convolution.